 \documentclass[preprint,5p,times,twocolumn]{elsarticle}

\usepackage[T1]{fontenc}

\usepackage{amsmath,amssymb,amsfonts}
\usepackage{algorithmic}
\usepackage{graphicx}
\usepackage{textcomp}
\usepackage{xcolor}
\usepackage{multirow}
\usepackage{booktabs}
\usepackage{subcaption}
\usepackage[separate-uncertainty=true]{siunitx}
\usepackage[colorinlistoftodos]{todonotes}
\usepackage{listings}
\usepackage{hyperref}

\journal{ }

\begin{document}

\begin{frontmatter}



\title{Deciphering boundary layer dynamics in high-Rayleigh-number convection using 3360 GPUs and a high-scaling in-situ workflow}

\author[jsc]{Mathis Bode\corref{cor1}}
\ead{m.bode@fz-juelich.de}
\cortext[cor1]{Corresponding author}
\author[jsc]{Damian Alvarez}
\author[uiuc,mcs]{Paul Fischer}
\author[eth]{Christos E. Frouzakis}
\author[jsc]{Jens Henrik G\"obbert}
\author[anl]{Joseph A. Insley}
\author[uiuc,mcs]{Yu-Hsiang Lan}
\author[anl,uic]{Victor A. Mateevitsi}
\author[mcs]{Misun Min}
\author[anl,uic]{Michael E. Papka}
\author[anl]{Silvio Rizzi}
\author[tui]{Roshan J. Samuel}
\author[tui,nyu]{J\"org Schumacher}


\affiliation[jsc]{organization={J\"ulich Supercomputing Centre, Forschungszentrum J\"ulich GmbH},
            city={52425 J\"ulich},
            country={Germany}}
            
\affiliation[uiuc]{organization={Siebel School of Computing and Data Science, University of Illinois at Urbana-Champaign},
            addressline={201 N Goodwin Ave},
            city={Urbana},
            postcode={61801},
           state={IL},
            country={USA}}

\affiliation[mcs]{organization={Mathematics and Computer Science Division, Argonne National Laboratory},
            addressline={9700 S Cass Ave},
            city={Lemont},
            postcode={60439},
           state={IL},
            country={USA}}

\affiliation[eth]{organization={CAPS Laboratory, Department of Mechanical and Process Engineering, ETH Z\"urich},
            city={8092 Z\"urich},
            country={Switzerland}}

\affiliation[anl]{organization={Argonne Leadership Computing Facility, Argonne National Laboratory},
            addressline={9700 S Cass Ave},
            city={Lemont},
            postcode={60439},
           state={IL},
            country={USA}}


 \affiliation[uic]{organization={Electronic Visualization Laboratory, University of Illinois Chicago},
            addressline={1200 W Harrison St},
            city={Chicago},
            postcode={60607},
           state={IL},
            country={USA}}

\affiliation[tui]{organization={Institute of Thermodynamics and Fluid Mechanics, Technische Universit\"at Ilmenau},
            addressline={P.O.Box 100565}, 
            city={98684 Ilmenau},
            country={Germany}}

\affiliation[nyu]{organization={Tandon School of Engineering, New York University},
            city={New York City},
            postcode={11201}, 
           state={NY},
            country={USA}}

\begin{abstract}
Turbulent heat and momentum transfer processes due to thermal convection cover many scales and are of great importance for several natural and technical flows. One consequence is that a fully resolved three-dimensional analysis of these turbulent transfers at high Rayleigh numbers, which includes the boundary layers, is possible only using supercomputers. The visualization of these dynamics poses an additional hurdle since the thermal and viscous boundary layers in thermal convection fluctuate strongly. In order to track these fluctuations continuously, data must be tapped at high frequency for visualization, which is difficult to achieve using conventional methods. This paper makes two main contributions in this context. First, it discusses the simulations of turbulent Rayleigh-B\'{e}nard convection up to Rayleigh numbers of $Ra=10^{12}$ computed with NekRS on GPUs. The largest simulation was run on 840 nodes with 3360 GPU on the JUWELS Booster supercomputer. Secondly, an in-situ workflow using ASCENT is presented, which was successfully used to visualize the high-frequency turbulent fluctuations.
\end{abstract}




\end{frontmatter}



\section{Introduction}

The availability of exascale supercomputers is both, a great opportunity to address unsolved scientific questions and advance technology, and a tremendous challenge for all scientists involved. Faster computers often result in even larger data sets, making it more difficult to process, visualize, and analyze that data. Other challenges include scalability issues of numerical algorithms, data security in increasingly heterogeneous or modular computing environments, and efficient use of complex high-performance computing (HPC) systems.

Solving the Navier-Stokes equations (NSEs) or simplified forms of fluid flow problems (collectively referred to as Computational Fluid Dynamics - CFD) is one of the traditional large application areas for supercomputers~\cite{Anderson:book:CFD,Yeung:PRF2020}. Due to the coupling of scales in turbulent flows, potential multi-physics effects, and the numerical difficulty of the problem, many fluid flow applications are still computationally too expensive even for exascale systems~\cite{Verma:SNC2020,Trebotich:JFE2024}.

NekRS is one of the frontrunner codes for solving low-Mach/incompressible NSEs on GPUs~\cite{Fischer:PC2022,NekRS:2023GBA} and is applied in this paper to decipher boundary layer effects in a three-dimensional (3D) flow subject to turbulent heat transfer at high Rayleigh numbers. For this purpose, a large-scale simulation with 840 nodes/3360 GPUs, i.e., 90\,\% of all JUWELS Booster nodes, was performed for almost 24 hours to reach a statistically steady state of turbulence plus another 6 hours utilizing in-situ visualization on JUWELS Booster. JUWELS Booster, hosted at the Jülich Supercomputing Centre (JSC), was ranked 18th on the TOP500 list~\cite{top500} when the presented simulations were performed. 

Turbulent heat transfer is a fundamental transport process in various fluid flows in nature and technology. The temperature difference in heat transfer problems causes density differences; denser fluid sinks due to gravity acceleration and lighter fluid rises. In the context of heat transfer in buoyancy-driven flows, also known as free (or natural) convection, the Rayleigh number is a central dimensionless parameter. The Rayleigh number $Ra$ relates the resulting buoyancy to viscous forces and thus describes the thermal driving of the fluid flow. Below a critical Rayleigh number $Ra_{\rm cr}$, heat transfer in an initially quiescent system takes place exclusively by heat conduction, i.e., without fluid motion. Above this level, $Ra>Ra_{\rm cr}$, heat is transferred by natural convection, i.e., conduction and additional fluid motion. The higher the Rayleigh number, the more turbulent the flow becomes as measured by the Reynolds number (which also increases roughly as $Re\sim Ra^{1/2}$. With increasing $Ra$ the amount of transfered heat increases and the smallest physically relevant scales in the flow decrease. Examples of turbulent heat transfer at high Rayleigh numbers are atmospheric flows, stellar convection or oceanic turbulence, where the relevant Rayleigh numbers are $Ra\gtrsim 10^{18}$. The paradigm to these complex heat transfer processes in nature and technology is the Rayleigh-B\'enard convection (RBC) case, a plane fluid layer between two infinitely extended parallel plates which is uniformly heated from below and cooled from above \cite{Ahlers:RMP2009,Chilla:EPJE2012}.

An infinitely extended natural convection layer is numerically modeled by a finite domain with a periodicity length $L$ in the horizontal directions and a layer height $H$. The aspect ratio of the domain follows to $\Gamma=L/H$ in the RBC setup (cf. \autoref{ssec:goveq}). A higher turbulence level (which is in line with a higher $Ra$) results in a scale ratio between the largest and smallest vortices or thermals in the flow which grows with $Ra^{3/2}$. With the largest length scale given as the constant domain size, this implies that as $Ra$ increases, the smallest relevant lengths become ever smaller. Thus, the grid resolution of the direct numerical simulation (DNS) must become correspondingly finer. For the analysis of RBC turbulence, highly accurate data are essential. With growing $Ra$, the local resolution must be increased to resolve all relevant effects. Ensuring that all data are sufficiently numerically resolved, a planar RBC setup with a square base was considered here for different $Ra$ and $\Gamma$. This was also achieved by analyzing which DNS conditions are reasonable in the context of computing costs, maximum system memory, and physical significance.

One major result of a recent series of DNS of RBC for $Ra\le 10^{11}$ was that the boundary layers of the temperature and velocity fields in a horizontally unconstrained configuration ($\Gamma=4$ and periodic boundary conditions in horizontal directions) are dominated by fluctuations \cite{Samuel2024}. This adds another layer of complexity from a HPC standpoint. For the largest case in this work ($Ra=\num{e12}$), the generated data (three velocity components, pressure, temperature) at each time step is 1.2\,TB in size or even 1.9\,TB if the spectral element grid is additionally written out. It is therefore not possible with classical visualization, i.e., by writing out complete checkpoint files and visualization as post-processing, to store data during a simulation in order to visualize the dynamics of the field fluctuations smoothly with respect to time. To fully visualize these fluctuations, which are physically essential to understand the local mechanisms of turbulent heat transfer, an in-situ workflow using ASCENT \cite{larsen2017alpine} was implemented and successfully applied on 840 JUWELS Booster nodes, as well as on other supercomputers at smaller scale. 

It has already been shown in the past that ASCENT can be used for large-scale simulations (up to \num{16000} GPUs~\cite{larsen_ascent_2022}). However, the use case considered here is different in that it requires both, large GPU numbers and a very high visualization frequency for the physical reasons mentioned above. As a consequence, the optimization of all sub-applications, i.e., NekRS and ASCENT, is essential, as otherwise the computational overhead will be too high. This paper presents such an optimized implementation and coupling.

In-situ visualization allows data visualization directly in the GPU memory. For this purpose, the simulation is periodically interrupted after a defined number of time steps and images are rendered. This approach significantly reduces the writing of data and enables the temporally complete visualization of even high-frequency simulations. Obvious disadvantages are the overhead caused by interrupting the simulation and the fact that what is to be visualized must be defined in advance\footnote{ASCENT supports the reading of a yaml file in each time step in order to change the scene to be visualized on-the-fly. This was also used to create the visualization setup in this paper. Despite this possibility of "interactive" adaptation, a general understanding of the simulation is often necessary in order to visualize in a meaningful way.}.

While JUWELS Booster has a peak performance of about 73 PetaFLOP/s, JSC will host the first European exascale system, JUPITER. JUPITER is currently expected to be available in 2025 and will have a peak performance of more than one ExaFLOP/s. As the software environment for JUPITER will be very similar to the current software environment used for JUWELS Booster, the presented results and especially the visualization discussions are relevant for future exascale approaches. The presented RBC case was also part of the JUPITER Benchmark Suite used for procuring JUPITER~\cite{herten2024jubench}. For the turbulent heat transfer case, the Rayleigh number $Ra$ is planed to be increased from \num{e12} to \num{e14} on JUPITER generating data that is more than 10 times larger while the storage bandwidth only increases by a factor of five. It can be assumed that the resulting fluctuations in the boundary layers will cover an even broader range including higher frequencies. Without in-situ methods, visualizations will then become even more challenging.

The focus of this paper is on HPC and related technical aspects. Therefore, the physical results of the large-scale simulation at $Ra=10^{12}$ are presented in brief, but not analyzed in depth. The main contributions of this paper are:
\begin{itemize}
    \item Analysis of the scaling properties of NekRS with respect to an RBC case;
    \item Presentation of a comprehensive RBC database with variations in $Ra$ and $\Gamma$;
    \item Execution and analysis of a DNS of RBCusing NekRS with Rayleigh number of \num{e12} on 3360 GPUs on JUWELS Booster;
    \item Implementation and successful application (optimization) of an ASCENT-based in-situ workflow at scale;
    \item Discussion of HPC properties also relevant to exascale computing.
\end{itemize}

\section{Background}
In the realm of modern HPC systems, the heterogeneity of computational resources is now the new norm, with CPU and GPU resources collaboratively tackling intricate scientific inquiries. For illustration, the forthcoming JUPITER system at JSC will be comprised of about \num{6000} compute nodes, each consisting of four NVIDIA GH2000 superchips, an innovative amalgamation of CPU and GPU technologies within a unified chipset. Similarly, the Argonne National Laboratory's Aurora system features \num{10624} nodes, each equipped with two Intel Xeon CPU Max Series processors alongside six Intel Data GPU Max Series GPUs. The Oak Ridge National Laboratory's Frontier supercomputer, with its \num{9402} nodes featuring one AMD EPYC CPU and four AMD Instinct MI250X GPUs, underscores the trend towards diverse computational architectures.

This paradigm shift towards heterogeneous computing environments has been pivotal in overcoming the exascale computing threshold, heralding a new epoch of scientific exploration. However, this advancement is not without its challenges. Primarily, science simulation codes, originally optimized for CPU-centric HPC architectures, necessitate significant refactoring to harness the computational prowess of GPUs. Furthermore, the heterogeneity of these systems introduces an additional layer of complexity, as each hardware vendor promotes a distinct programming model—CUDA for NVIDIA, SYCL for Intel, and HIP for AMD—complicating the development and optimization of portable, efficient scientific applications.

\subsection{HPC Systems}
This work relied primarily on JUWELS Booster for its large-scale simulation at $Ra=10^{12}$ and measurements. The presented production results are among the largest simulations ever performed on JUWELS Booster. The visualization results were partly reproduced on three supercomputers with NVIDIA GPUs, Leonardo Booster at CINECA, MareNostrum 5 ACC at Barcelona Supercomputing Centre, and Polaris at Argonne National Laboratory, to demonstrate the flexibility of the workflow. This emphasizes the general nature of the presented large-scale visualization solution.

\subsubsection{JUWELS Booster}
JUWELS Booster, a GPU accelerated supercomputing module was deployed in 2020. The module is integrated in the larger JUWELS system, whose first module -- now known as JUWELS Cluster -- was deployed in 2018. This GPU accelerated module comprises 936 nodes, is fully directly liquid cooled, and is based on Atos' Sequana XH2000 architecture. At deployment time it ranked \#7 in the TOP500 list, and \#3 in the GREEN500 list. JUWELS is the current flagship system at JSC, and the blueprint for the upcoming Exascale-class JUPITER system, scheduled to be deployed in 2025.

\begin{figure}
\centerline{\includegraphics[width=0.50\textwidth]{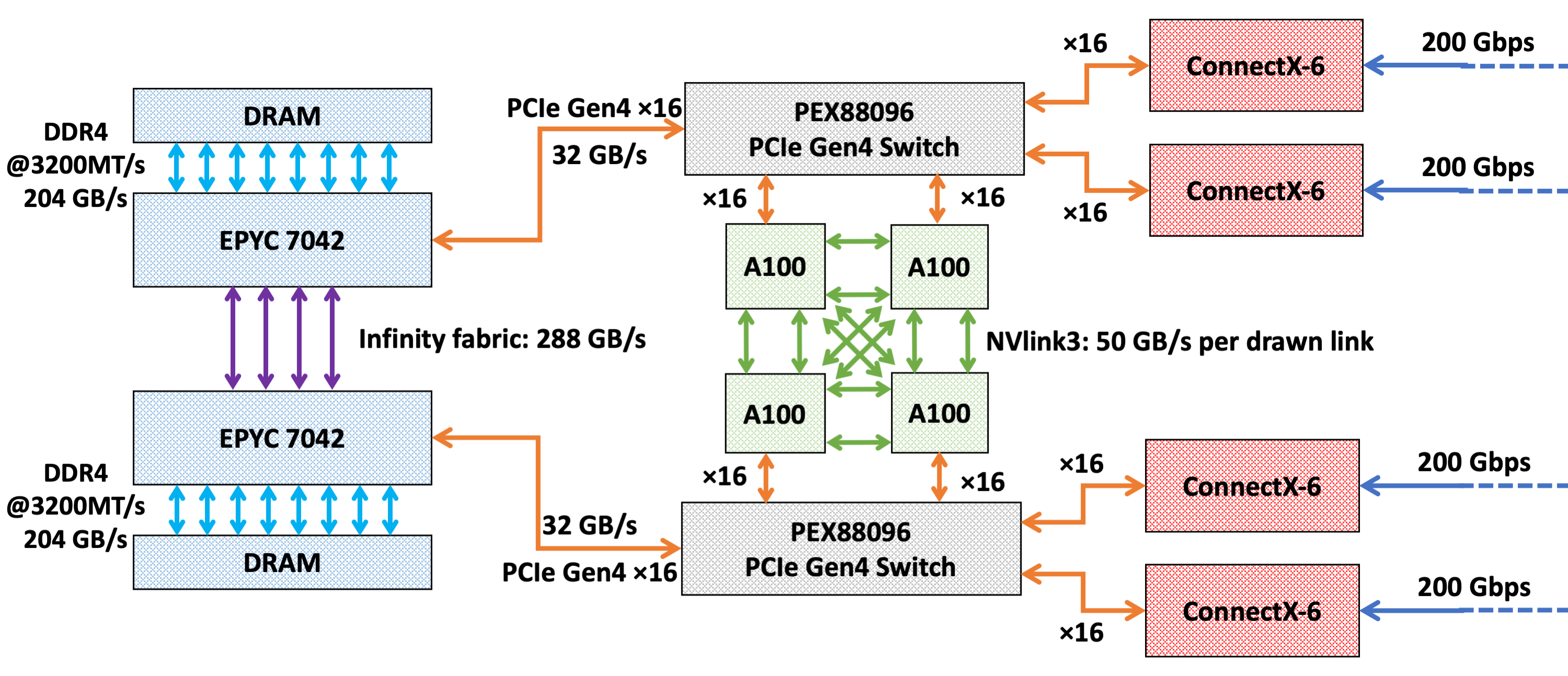}}
  \caption{
  JUWELS Booster node architecture.
  }
\label{fig:booster_node}
\end{figure}

JUWELS Booster's node architecture shown in \autoref{fig:booster_node} relies on 2x AMD EPYC 7042 processors that orchestrate 4x NVIDIA A100 GPUs. These GPUs have the SXM4 form factor, and are integrated in a Redstone board, in which the GPUs are interconnected in direct all-to-all NVLink3 fabric. The total NVLink bandwidth per GPU is 300\,GB/s, with each of the links capable of full-duplex communication, resulting in 600 GB/s of bidirectional bandwidth. For internode communication, each GPU has an HCA to an HDR InfiniBand fabric. The injection bandwidth per node is 800\,Gbps.

The InfiniBand fabric in JUWELS is the result of integrating 3 different fabrics: A Fat Tree on the JUWELS Cluster module, a small Fat Tree as part of a flash storage based on IME, and a DragonFly+ fabric with adaptive routing on the JUWELS Booster module. The JUWELS Booster module contains a total of 20 DragonFly+ groups, each implementing a Fat Tree within the group. Inside the group, copper cables are used, resulting in a very cost effective, scalable and performant fabric that includes almost \num{700} switches, \num{35000} ports and \num{6500} HCAs.

Cluster and Booster modules of JUWELS are described in greater detail in~\cite{juwels}. Given the similarities between JUWELS Booster, and the upcoming Exascale-capable JUPITER Booster module, it is expected that the workflow presented here is generally transferable. Furthermore, NekRS was part of the JUPITER benchmark suite~\cite{herten2024jubench} that was used for procuring JUPITER, so that it should perform well on JUPITER once it is ready for production.

\subsubsection{Other Supercomputers}
In addition to JUWELS Booster, the simulations including the visualization were tested on other supercomputers with different GPUs or network configurations to ensure that the workflow works independently of the system used. This is particularly important if large-scale data sets are to be analyzed or further processed on other supercomputers; an increasingly relevant workflow for exascale simulations. For these tests, the scope was limited to systems with NVIDIA GPUs for the sake of simplicity. In principle, however, the workflow should also be portable on AMD and Intel GPUs. All supercomputers relevant for this work are briefly compared in \autoref{tab:supercomputers}.
\begin{table*}[h]
\caption{Overview of supercomputers. JUPITER's specifications are projected.}
\centering
\begin{tabular}{l c c c c c}
\toprule
& \textbf{JUWELS} & \textbf{JUPITER} & \textbf{Leonardo} & \textbf{MareNostrum} & \textbf{Polaris}\\
& \textbf{Booster} & \textbf{Booster} & \textbf{Booster} & \textbf{5 ACC} & \\
\midrule
Number of nodes & \num{936} & \num{5884} & \num{3456} & \num{1120} & \num{560} \\
Number of GPUs & \num{3744} & \num{23536} & \num{13824} & \num{4480} & \num{2240} \\
GPU & NVIDIA & NVIDIA & NVIDIA & NVIDIA & NVIDIA \\
& A100 & H100 & A100 & H100 & A100 \\
& (HBM2, 40\,GB) & (HBM3, 96\,GB) & (HBM2e, 64\,GB) & (HBM2e, 64\,GB) & (HBM2, 40\,GB) \\
Total GPU memory (TB) & \num{150} & \num{2259} & \num{885} & \num{287} & \num{90} \\
Node connectivity & $4\times$HDR200 & $4\times$NDR200 & $2\times$HDR200 & $4\times$NDR200 & $2\times$Slingshot 11 \\
& InfiniBand & InfiniBand & InfiniBand & InfiniBand & \\
\bottomrule
\end{tabular}
\label{tab:supercomputers}
\end{table*}



\subsection{NekRS}
NekRS employs the spectral element method (SEM) \cite{patera1984spectral}, a high-order weighted residual approach that combines fast matrix-free tensor-product operator evaluation with rapidly-convergent nodal-based Gauss-Lobatto-Legendre (GLL) quadrature.   The spectral element discretization consists of $N_e$ elements of order $p$, for a total number of unknowns $n \approx N_e p^3$ per scalar field.  The tensor-product ($ijk$) ordering of the unknowns in the canonical reference element, $(r,s,t) \in {\hat \Omega}=[-1,1]^3$, allows operators to be expressed as tensor contractions leading to respective storage and computational complexities $O(n)=O(N_e p^3)$ and $O(np)=O(N_e p^4)$, respectively \cite{deville:book2002}, which is a significant reduction compared to the $O(N_e p^6)$ complexity of the traditional $p$-type finite element method.  Moreover, the leading order work terms are efficient dense tensor contractions of $(p+1) \times (p+1)$ operator matrices, identical for all elements, applied to the spectral element data.  As noted early on by Orszag \cite{orszag1979spectral}, the fast tensor-product forms readily carry over from the reference element to deformed elements in physical space through application of the chain rule.

Realization of reasonable (say, 80\%) parallel efficiency on modern nodes, such as with NVIDIA A100 GPUs and HDR200 InfiniBand, requires about 2--5M grid points per MPI rank \cite{min2022optimization,min22c}. For polynomial order $p=7$, this requirement implies having $\approx\num{8000}$ elements per rank. In this case, there are enough interior elements on each MPI rank to allow overlapping computation and communication. One begins by updating residuals on the surface elements and initiates the surface-element data exchange. Next, interior element residuals are updated and exchanged within each rank. Finally, the incoming communication data is used to complete the residual values on the subdomain surface. The stiffest substep in time advancement of the NSE is the pressure Poisson solve that enforces incompressibility at each time step. NekRS supports a variety of multilevel preconditioners for this problem, with the default corresponding to $p$-multigrid that uses 4th-kind Chebyshev-acceleration of overlapping Schwarz smoothers \cite{phillips2023spectral}. To save on memory bandwidth, the majority of the preconditioner steps are implemented in FP32.
 
NekRS is mainly written in C++ and the kernels are implemented using the portable Open Concurrent Compute Abstraction (OCCA) library \cite{medina2014occa, medina2015okl}, in order to abstract between different parallel languages. In this way, the MPI+X hybrid parallelism is supported seamlessly across different backends such as CUDA, HIP, OPENCL, and DPC++, as well as CPUs. On each node, one MPI rank per GPU is used and all data resides on the device, with a copy back to the host only when needed (e.\,g., for I/O or data analysis). OCCA enables the implementation of the parallel kernel codes in the slightly decorated C++ language OKL \cite{medina2015okl} and translates the OKL source codes into the desired language. The compute intensive kernels are benchmarked into several pre-optimized versions and the just-in-time compilation autotunes the kernels at runtime to pick the fastest version based on the problem size and the target architecture. Details of the used version are given in \autoref{tab:software-release}.
\begin{table}[h]
\caption{Software versions and release dates}
\centering
\begin{tabular}{lll}
\toprule
\textbf{Software} & \textbf{Version} & \textbf{Release date} \\
\midrule
GCC       & 12.3.0         & 08-May-2023   \\
OpenMPI   & 4.1.5          & 23-Feb-2023   \\
CUDA      & 12.2           & 31-Jul-2023   \\
VTK-m     & 2.1.0          & 29-Nov-2023   \\
ASCENT    & commit 6d1ca3f & 12-Jan-2024   \\
NekRS     & v23.0          & 30-May-2023   \\
          & + ASCENT coupling & Mar-2024   \\
\bottomrule
\end{tabular}
\label{tab:software-release}
\end{table}

\subsection{ASCENT}
As the exascale era starts, the capability to conduct simulations at scales previously unimaginable is within reach. This increase in computational power is not without its challenges, particularly in the domain of data I/O. The evolution of computing power has outpaced the advancements in I/O technologies, leading to a significant bottleneck: saving high-frequency simulation results to storage systems can drastically impede the overall simulation performance. This discrepancy threatens to undermine the efficiency and feasibility of running large-scale simulations, a critical concern for the HPC community.

A strategic solution to this challenge lies in the adoption of in-situ analysis and visualization libraries. These tools offer a paradigm shift by enabling the analysis and rendering of visualizations directly from data resident in memory, circumventing the need for costly data movement to and from storage devices. The core advantage of this approach is the substantial reduction in I/O overhead, thus allowing scientists to interact with and analyze simulation data at much higher frequencies than would be possible if traditional I/O processes were employed. By facilitating access to high-fidelity, high-frequency data, in-situ technologies enable researchers to capture transient phenomena and subtle interactions that may otherwise be overlooked, reducing the risk of missing critical discoveries that could occur during the intervals between conventional I/O operations.

Among the forefront of in-situ analysis and visualization libraries tailored for HPC simulations, ASCENT \cite{larsen_ascent_2022}, Catalyst2 \cite{catalyst_revised2021}, LibSim \cite{kuhlen_parallel_2011, childs_situ_2012}, and SENSEI \cite{ayachit_sensei_2016} are particularly noteworthy. ASCENT, a component of the ALPINE initiative and supported by the U.S. Department of Energy's Exascale Computing Project (ECP) \cite{messina_exascale_2017}, distinguishes itself as a lightweight library with minimal external dependencies, leveraging VTK-m \cite{moreland_vtk-m_2016} for its rendering capabilities. Catalyst, developed by Kitware, builds upon VTK to offer advanced visualization workflows, demonstrating the library’s robustness in handling complex visualization tasks. Similarly, LibSim, designed to work seamlessly with VisIt \cite{childs_visit_2012}, further exemplifies the integration of versatile visualization tools in supporting HPC simulations.

In selecting an appropriate in-situ analysis and visualization library for this work, ASCENT emerged as the preferred choice, attributable to several distinct advantages. Primarily, its lightweight design and ease of integration into existing simulation code bases significantly reduce the implementation overhead. Furthermore, ASCENT supports zero-copy GPU-to-GPU data passing, which enable direct transfer of device pointers between the simulation code and ASCENT, ensuring that data remains exclusively on the GPU. Such an approach effectively eliminates the need for the data to traverse back to the CPU, addressing a common bottleneck in high-performance computing applications.

The implications of this GPU-centric data handling strategy are profound, notably in terms of minimizing the memory footprint of the in-situ visualization process. Traditional data movement to the CPU incurs significant memory overhead and duplication \cite{mateevitsi2023scaling}, and by circumventing this transfer, ASCENT facilitates a more efficient utilization of computational resources.

\section{ASCENT-NekRS Coupling}
This section summarizes the ASCENT-NekRS coupling that is an essential part for performing the fully time-resolved visualizations in this work.

\subsection{Implementation}
Given that the data is already on the GPU, transferring it to the CPU would be inefficient due to the overhead involved. Furthermore, depending on the renderer configuration, the data might be processed on the CPU or GPU, with the former being slower than the latter. Our instrumentation of NekRS with Ascent retains the data on the GPU, passing only a device pointer to ASCENT, as demonstrated in Listing~\ref{listing:conduit}. ASCENT utilizes Conduit \cite{harrison2022conduit} for data description.

\begin{lstlisting}[language=C++,breaklines,caption=Example of passing the pressure values as a device pointer to ASCENT, label=listing:conduit]
conduit::Node mesh_data;
mesh_data["fields/pressure/association"]  = "vertex";
mesh_data["fields/pressure/topology"]     = "mesh";
mesh_data["fields/pressure/values"].set_external( (dfloat*) nrs->o_P.ptr(), mesh->Nlocal);
\end{lstlisting}

The instrumentation passes the following to ASCENT: the mesh, the velocity vector, the pressure, temperature, and any other user-defined scalar field. Additionally, the mesh connectivity array is calculated and sent to ASCENT, which is the only data residing in host memory.

Initially, the Conduit node that holds all the pointers to the data is populated once at the simulation's initialization. Since the pointers do not change from time step to time step, these values remain constant. ASCENT is then called at every time step where analysis is requested as shown in Listing~\ref{listing:ascent_update}.

\begin{lstlisting}[language=C++, caption=Code that gets called every time ascent analysis is needed, label=listing:ascent_update]
// Update timestep numbers
mesh_data["state/cycle"] = tstep;
mesh_data["state/time"] = time;

// call ascent
ascent.publish(mesh_data);
conduit::Node actions;
ascent.execute(actions);
\end{lstlisting}

Finally, ASCENT needs to exit gracefully and delete all previously allocated memory, so \verb|ascent.close();| must be called before exiting the simulation.

With the above instrumentation, NekRS now leverages the full power of ASCENT, including HDF5 I/O support, in transit with ADIOS2, and in situ rendering with \verb|vtk-m|. Users can perform in-situ and in-transit pipelines by modifying the \verb|ascent_actions.yaml| file.

\subsection{Further Improvements}
Initial runs demonstrated that the time per time step was approximately 20\,\% slower with the ASCENT instrumentation compared to without it. Further investigation revealed that various background checks occur when calling \verb|ascent.execute(actions);| even if no in-situ or in-transit pipeline is triggered. To avoid this overhead, ASCENT is called only when in-situ or in-transit workflows are needed, or when its trigger mechanism needs to check whether to run its pipelines.

\subsection{Technical Challenges}
The usage of ASCENT with NekRS needs to address two technical challenges:
\begin{itemize}
\item Data Location Awareness in ASCENT: A primary challenge involved the data transfer mechanism within ASCENT, particularly when interfacing with different extractors. ASCENT inherently passes data between filters and extractors without altering its location, necessitating explicit management of data residency. For instance, providing a GPU device pointer to ASCENT and subsequently invoking an HDF5 extract operation led to application crash. This issue stems from the HDF5 extractor's expectation for data to reside on the CPU, without the capability to autonomously transfer the data from the GPU. To mitigate this, we manually moved the data to the CPU prior to the extraction processes, highlighting the importance of data residency awareness when leveraging extraction capabilities of ASCENT (cf. \autoref{ssec:fur}).

\item Bug in OCCA's Slice Operation: Another technical hurdle was identified in the slice operation provided by OCCA, as integrated within NekRS v23, which returned incorrect data. The solution involved bypassing the faulty OCCA slice operation in favor of direct manipulation with C++ pointers. The bug was fixed on the NekRS next branch.
\end{itemize}

\section{Rayleigh-B\'enard convection model and dynamics}
\begin{figure}
\centerline{\includegraphics[width=0.50\textwidth]{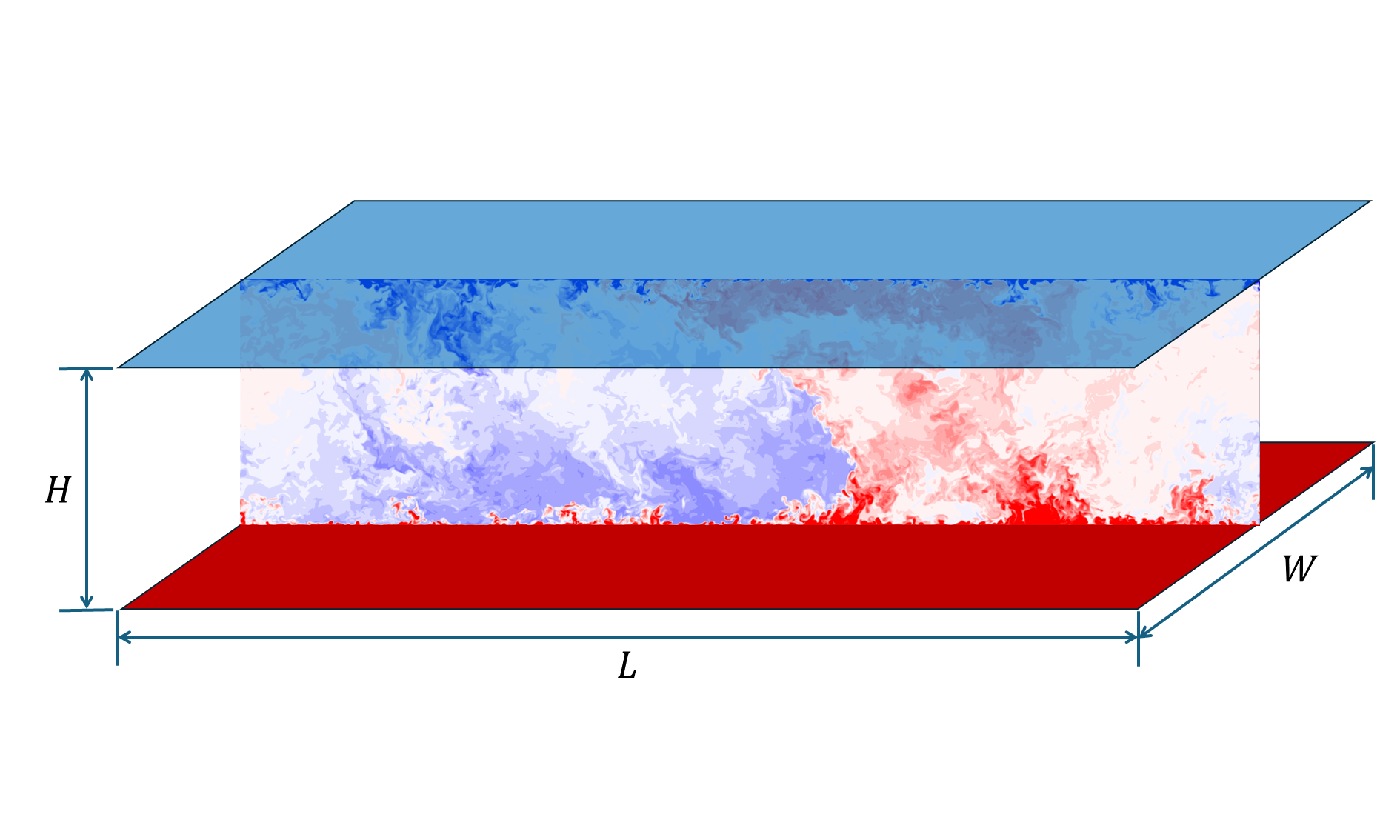}}
  \caption{
  Schematic of Rayleigh-B\'enard convection setup. The top plate is held at a uniform temperature $T=T_t$, the bottom plate at $T=T_b>T_t$.  
  }
\label{fig:rbc_setup}
\end{figure}

The RBC setup is a canonical case of fluid dynamics. It is used to understand the physics of thermal convection, which is commonly encountered in cooling systems, atmospheric flows, and astrophysical phenomena as already mentioned in the introduction. \autoref{fig:rbc_setup} shows a schematic of the RBC setup with the hot bottom and cold top plates, colored red and blue respectively. A representative structure of the flow on a vertical slice is also shown.
The RBC simulation domain comprises a horizontal layer of fluid confined between a pair of infinitely extended parallel plates placed at a vertical distance $H$. Periodic boundary conditions are applied in the other two directions. Slabs with a quadratic horizontal cross section are used, i.e., $L=W$ in \autoref{fig:rbc_setup}, to ensure the same level of “unboundedness” in both periodic directions. The bottom (hot) plate is maintained at a higher temperature than the top (cold) plate, so that the warmer fluid at the bottom rises due to buoyancy forces, while the cooler fluid at the top flows downward to replace it, thus establishing the convective flow~\cite{Ahlers:RMP2009,Chilla:EPJE2012,Getling:book,Verma:book:BDF}.

\subsection{Governing Boussinesq equations}
\label{ssec:goveq}
Numerical simulations of RBC are typically performed by solving the three-dimensional Navier-Stokes equations, along with the temperature and continuity equations,
\begin{equation}
    \frac{\partial \mathbf{u}}{\partial t} + \mathbf{u} \cdot \nabla\mathbf{u} = -\nabla p + T \hat{z} + \sqrt{\frac{Pr}{Ra}} \nabla^2 \mathbf{u},
    \label{eq:nse_vel}
\end{equation}
\begin{equation}
    \frac{\partial T}{\partial t} + \mathbf{u} \cdot \nabla T = \frac{1}{\sqrt{Ra Pr}} \nabla^2 T,
    \label{eq:nse_tmp}
\end{equation}
\begin{equation}
    \nabla \cdot \mathbf{u} = 0.
    \label{eq:nse_cnt}
\end{equation}
Here, $\mathbf{u}$, $p$, and $T$ are the velocity, pressure and temperature fields, respectively. This set of equations is also known as the Boussinesq approximation of natural convection. The term $T \hat{z}$ in Eq.~\eqref{eq:nse_vel} is the body force arising from buoyancy, which drives the convective flow. Since this force acts along the direction of gravity, it contributes only to the vertical $\hat{z}$ component of momentum.The equations are non-dimensional and depend on the dimensionless numbers, Rayleigh number ($Ra$), and Prandtl number ($Pr$) to adjust the fluid properties.

Rayleigh number is the ratio between the strength of buoyancy and dissipation, $Ra = \alpha g \Delta H^3 / (\kappa \nu)$, where $\alpha$ is the bulk thermal expansion coefficient, $g$ is the acceleration due to gravity, $\Delta = T_b - T_t$ is the difference in temperature between the bottom ($T_b$) and top ($T_t$) plates, $H$ is the vertical height between the plates, $\nu$ and $\kappa$ are the kinematic viscosity and thermal diffusivity respectively. The Prandtl number is the ratio between kinematic viscosity and thermal diffusivity, $Pr = \nu/\kappa$. Finally, the aspect ratio of the domain $\Gamma = L/H$ is the third non-dimensional parameter.

The key factor that determines the resolution of the grid is the boundary layer (BL) that develops at the top and bottom plates. In NekRS, the domain is first discretized into \(N_{e}\) hexahedral elements, then the element-wise data is represented by 3D tensor of $p$th order polynomials. That is $N_{e}p^3$ grid points in total. For a finer resolution, either the number of elements or the polynomial order (for NekRS $p\le 10$) can be increased. The number of vertical collocation points within the boundary layer (BL) is denoted by $N_{BL}$, and ideally $N_{BL} \geq 10$ to properly resolve the flow within the BL as discussed below. As $Ra$ increases, the thickness of the thermal  BL (denoted by $\delta_T$) decreases, which places a lower bound on the vertical resolution, and correspondingly the total number of points.

\subsection{Resolution requirements}
A necessary first step for implementing physically correct DNS is understanding the resolution needed. In general, larger $Ra$ require higher resolution because the resulting flow becomes more turbulent, i.e., the ratio of largest to smallest scales increases. Since higher resolution is in line with higher memory requirements on the GPUs, and GPU memory is the physical limit with regard to the maximum achievable $Ra$ on each supercomputer (simulation runtimes are “only days” and therefore manageable), a resolution that is just sufficient should be chosen to achieve the highest possible $Ra$.

\begin{figure}[h!]
\centerline{\includegraphics[width=0.45\textwidth]{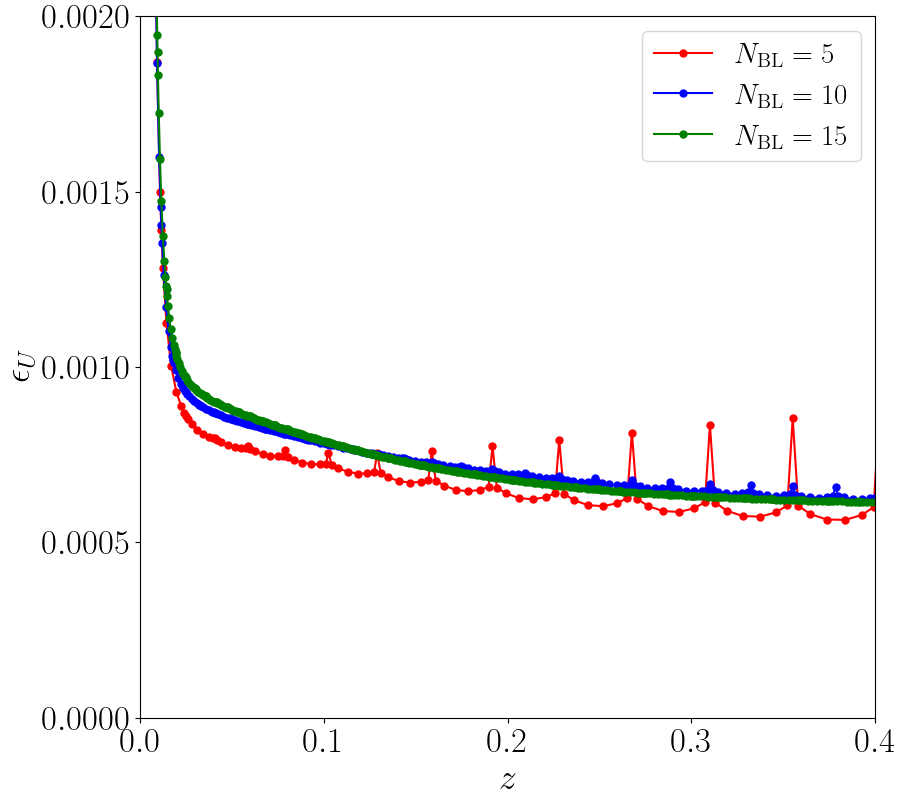}}
  \caption{
  Kinetic energy dissipation as a function of the distance from the bottom boundary for $N_{BL}=5,10,15$ and $Ra=\num{e9}$.
  }
\label{fig:resolution}
\end{figure}

Scheel et al.~\cite{scheel2013} studied the impact of the spatial descritization approach on RBC results in a closed cylindrical cell. Following their analysis, \autoref{fig:resolution} shows the kinetic energy dissipation rate $\epsilon_U$, defined as
\begin{equation}
\epsilon_U = 2\nu \mathbf{S}:\mathbf{S},    
\end{equation}
as a function of the distance from the bottom plate for three boundary layer resolutions; $\mathbf{S}=\frac{1}{2}(\nabla\mathbf{u}+(\nabla\mathbf{u})^\intercal)$ is the symmetric strain-rate tensor. It is clearly visible that $N_{BL}=5$ is significantly underresolved. The results for $N_{BL}=10$ and $N_{BL}=15$ are not identical, but they clearly approach each other.

Based on these results, the GPU memory requirements for different RBC setups can be estimated. This is summarized in \autoref{tab:memory} for two different boundary layer resolutions $N_{BL}$ with $p=7$ as the order of the spectral elements. Furthermore, the grid stretching function also plays a critical role in determining the resolution of the boundary layers. The tangent-hyperbolic stretching function commonly used in literature is applied along the vertical direction while generating the element map. The stretching parameter for this function varies from 1.3 to 2.0, with the highest stretching values used for grids at the highest $Ra$ to fit as many points as possible in the thermal boundary layer.
\begin{table*}[h]
\caption{Estimation of GPU memory requirement for RBC simulations with different $Ra$ and $\Gamma$. All memory estimates assume $p=7$ are and are given in TB. The grid stretching varies between $1.3$ and $2.0$ with higher stretching values for higher $Ra$.}
\centering
\begin{tabular}{lcccccc}
\toprule
& \multicolumn{6}{c}{\textbf{$\Gamma$}} \\
\textbf{$Ra$} & \multicolumn{2}{c}{$2$} & \multicolumn{2}{c}{$4$} & \multicolumn{2}{c}{$8$} \\
& $N_{BL}=10$ & $N_{BL}=15$ & $N_{BL}=10$ & $N_{BL}=15$ & $N_{BL}=10$ & $N_{BL}=15$ \\
\midrule
\num{E+05} & \num{4.57E-04} & \num{1.19E-03} & \num{1.83E-03} & \num{5.20E-03} & \num{7.32E-03} & \num{2.17E-02} \\
\num{E+06} & \num{1.54E-03} & \num{4.39E-03} & \num{6.17E-03} & \num{1.86E-02} & \num{2.47E-02} & \num{7.65E-02} \\
\num{E+07} & \num{9.08E-03} & \num{3.02E-02} & \num{3.63E-02} & \num{1.21E-01} & \num{1.45E-01} & \num{4.82E-01} \\
\num{E+08} & \num{4.28E-02} & \num{1.89E-01} & \num{1.76E-01} & \num{7.70E-01} & \num{7.03E-01} & \num{3.11E+00} \\
\num{E+09} & \num{2.73E-01} & \num{9.66E-01} & \num{1.11E+00} & \num{3.86E+00} & \num{4.43E+00} & \num{1.55E+01} \\
\num{E+10} & \num{1.14E+00} & \num{5.30E+00} & \num{4.58E+00} & \num{2.12E+01} & \num{1.83E+01} & \num{8.47E+01} \\
\num{E+11} & \num{3.60E+00} & \num{1.32E+01} & \num{1.45E+01} & \num{5.28E+01} & \num{5.82E+01} & \num{2.12E+02} \\
\num{E+12} & \num{2.67E+01} & \num{1.12E+02} & \num{1.07E+02} & \num{4.50E+02} & \num{4.29E+02} & \num{1.80E+03} \\
\num{E+13} & \num{1.45E+02} & \num{5.31E+02} & \num{5.82E+02} & \num{2.13E+03} & \num{2.33E+03} & \num{8.51E+03} \\
\num{E+14} & \num{4.57E+02} & \num{1.51E+03} & \num{1.83E+03} & \num{6.04E+03} & \num{7.31E+03} & \num{2.42E+04} \\
\num{E+15} & \num{3.22E+03} & \num{1.21E+04} & \num{1.29E+04} & \num{4.83E+04} & \num{5.16E+04} & \num{1.93E+05} \\
\bottomrule
\end{tabular}
\label{tab:memory}
\end{table*}

\begin{figure}[h!]
\centerline{\includegraphics[width=0.45\textwidth]{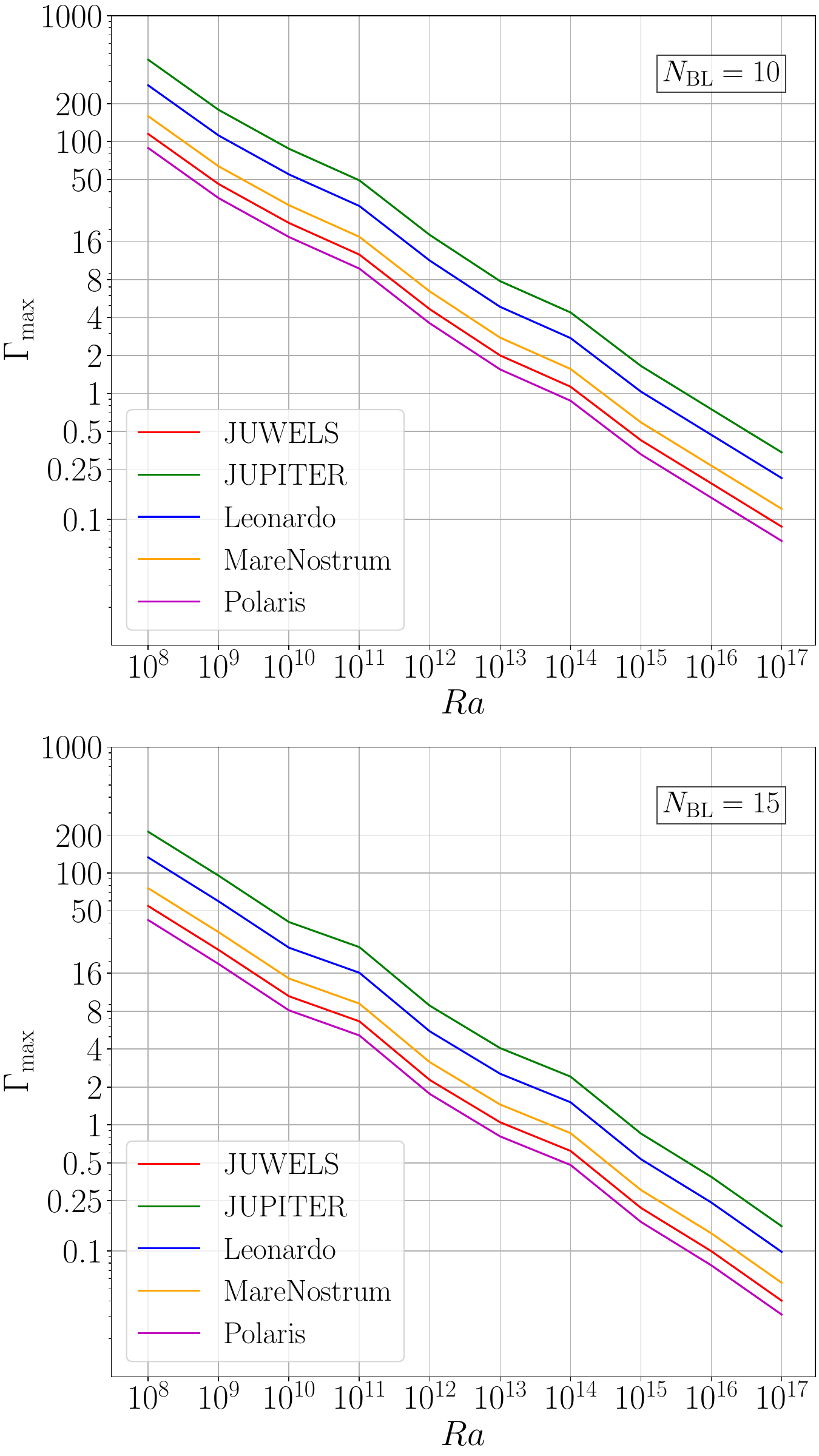}}
  \caption{
  Accessible configurations for different supercomputers for $N_{BL}=10$ (top) and $N_{BL}=15$ (bottom). All estimates assume $p=7$.
  }
\label{fig:area}
\end{figure}

The results from \autoref{tab:memory} are plotted in \autoref{fig:area} with their significance for five different supercomputers. The log-log plot illustrates two aspects. First, on JUWELS Booster the maximum accessible $Ra$ for $\Gamma=4$ is \num{e12} at a resolution of $N_{BL}=10$. Second, with the new exascale supercomputer JUPITER, the accessible parameter range shifts by two orders of magnitude to $Ra=\num{e14}$ for $N_{BL}=10$. 

\subsection{Choice of aspect ratio $\Gamma$}
\begin{figure}[h!]
\centerline{\includegraphics[width=0.50\textwidth]{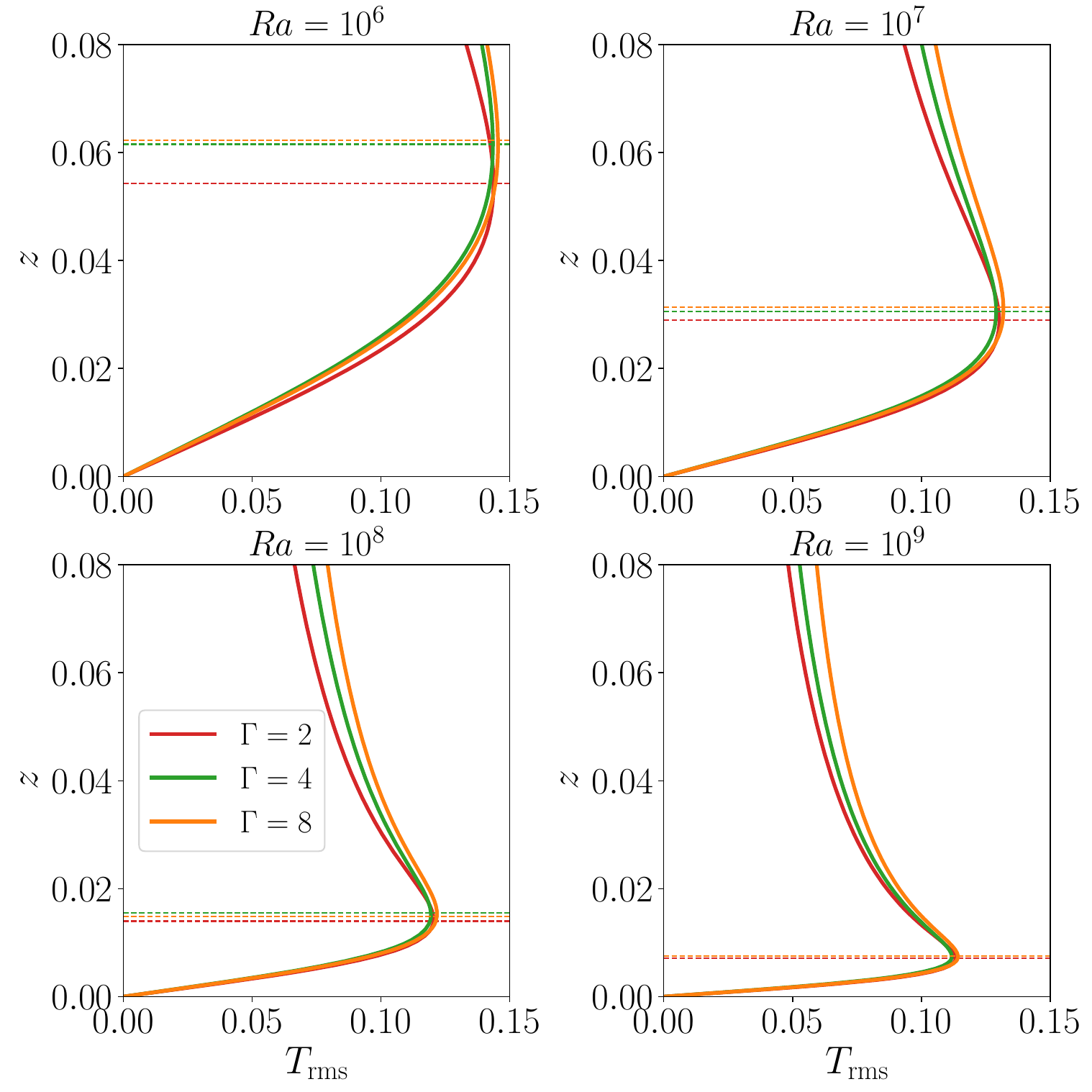}}
  \caption{
  Aspect ratio-dependence of RMS profiles of temperature
  }
\label{fig:profiles_trms}
\end{figure}

\begin{figure}[h!]
\centerline{\includegraphics[width=0.50\textwidth]{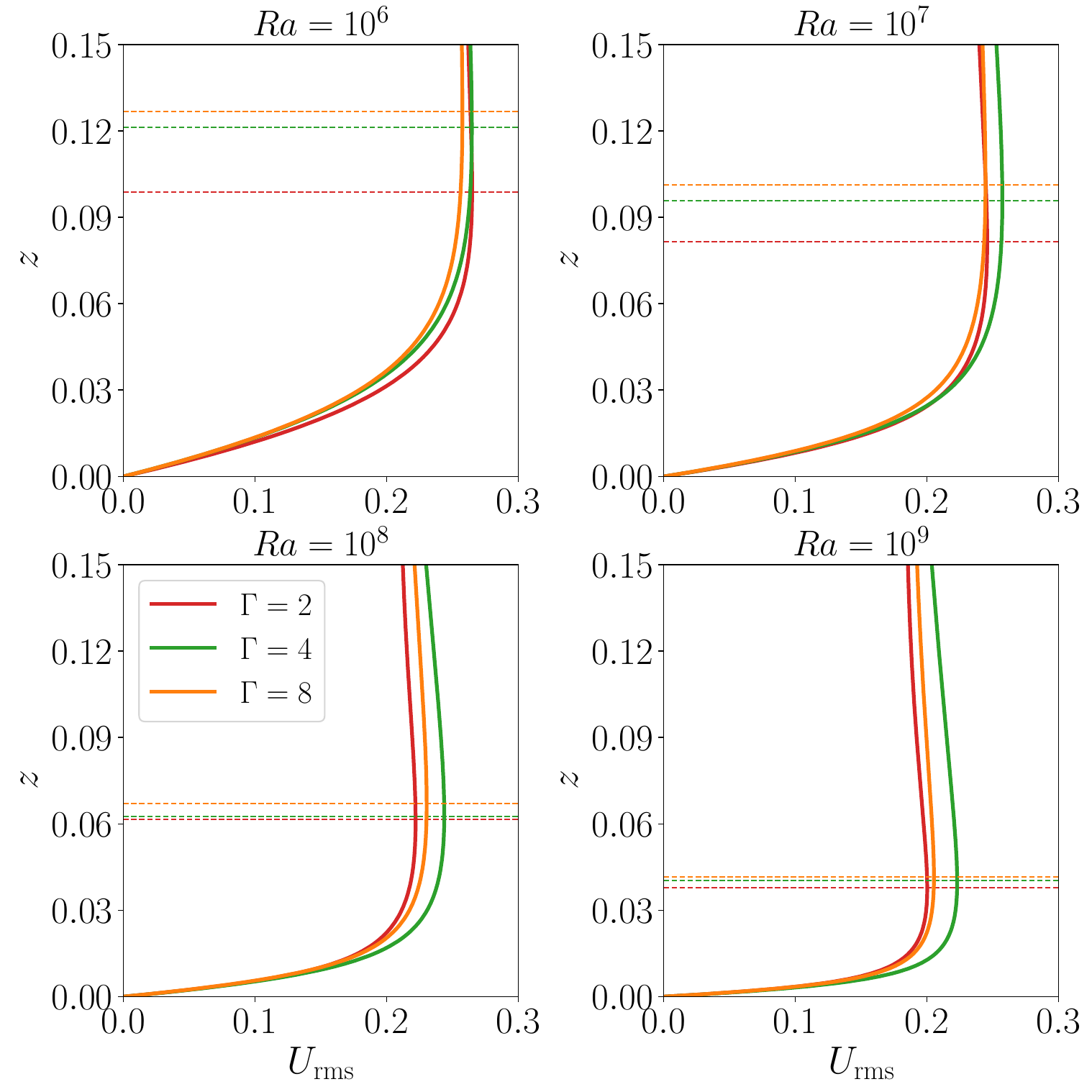}}
  \caption{
  Aspect ratio-dependence of RMS profiles of velocity
  }
\label{fig:profiles_urms}
\end{figure}

In addition to the resolution, the maximum possible $Ra$ is influenced by the choice of $\Gamma$. On the one hand, the GPU memory requirement (and the required computing time) is proportional to $\Gamma^2$. On the other hand, $\Gamma$ must be large enough for the flow to behave like that of infinite plates. The appropriate choice of $\Gamma$ is therefore essential. To understand the effect of $\Gamma$ on the flow, simulations were carried out for $Ra=\num{e6}, \num{e7}, \num{e8}, \num{e9}$ with $\Gamma=2,4,8$ and compared with regard to the resulting boundary layer heights. \autoref{fig:profiles_trms} compares the root mean square (RMS) profiles of the temperature and the resulting thermal boundary layer height. It can be seen that the thermal boundary layer height for $\Gamma=2$ is lowest for all $Ra$ and, in particular, is further away from the resulting boundary layer heights for $\Gamma=4$ and $\Gamma=8$ than the distance between them. Overall, the resulting thermal boundary layer height seems to converge between $\Gamma=4$ and $\Gamma=8$. For the profiles obtained here, the thermal boundary layer height for $\Gamma=4$ is slightly higher in one case, slightly lower in two cases, and almost identical to that for $\Gamma=8$ in one case.
The converging trend cannot be confirmed for the kinetic boundary layer height, which is determined using $U_{rms}$ as shown in \autoref{fig:profiles_urms}.

\subsection{Dataset for $\Gamma=4$}
\label{ssec:case}
The RBC setup with $\Gamma=4$ and $N_{BL} \ge 10$ was chosen as the best compromise for the plane-layer RBC studies at the highest possible $Ra$ in the context of this work. Summarizing, the DNS of RBC were performed in a Cartesian box of aspect ratio $\Gamma=4$, with periodic boundary conditions along the horizontal $x$ and $y$ directions. The top and bottom plates are isothermal and no-slip (which implies that all three velocity components are exactly zero at the walls).

The list of simulated cases along with their grid resolutions is provided in \autoref{tab:sim_details}. The table reports the Rayleigh number $Ra$, the number of spectral elements $N_e$, the polynomial order $p$ on each element in each space direction, the resulting unique gridpoints ($N_e \times p^3$), the number of grid points within the thermal boundary layer $N_{BL}$, an average time step size that can be advanced with each simulation step $dt$, and the total averaging time in free-fall units $\Delta t$. Almost 46.7 billion gridpoints are required at $Ra=10^{12}$. Furthermore, \autoref{tab:sim_details} also provides average time step sizes for time steps without and with visualization. These values are discussed in \autoref{sec:performance}.
\begin{table*}[t]
  \caption{Details of the simulations with $\Gamma=4$}
  \begin{center}
    \begin{tabular}{lccccccccc}
    \hline
    $Ra$        &       $N_e$                    & $p$  & gridpoints & \(N_{BL}\) & \(dt\)    & \(\Delta t\) & \(N_{nodes}\) & \(t_{novis} (s)\) & \(t_{vis} (s)\) \\
    \hline
    \(10^{5}\)  & \(100 \times 100 \times 64\)   & 5  & 80 million & 71  &  $1.33 \times 10^{-2}$  & 1000&6& \num{0.11+-0.003}           & \num{8.95+-0.051}          \\
    \(10^{6}\)  & \(100 \times 100 \times 64\)   & 7  & 220 million & 57 &  $5.73 \times 10^{-3}$  & 1000&12& \num{0.13+-0.005}           & \num{9.02+-0.029}                   \\
    \(10^{7}\)  & \(100 \times 100 \times 64\)   & 9  & 467 million & 42 &  $2.45 \times 10^{-3}$ & 1000&25& \num{0.15+-0.003}           & \num{9.78+-0.028}                   \\
    \(10^{8}\)  & \(150 \times 150 \times 96\)   & 7  & 740 million & 24    &  $1.82 \times 10^{-3}$  & 1000&40& \num{0.13+-0.004}           & \num{9.29+-0.053}                   \\
    \(10^{9}\)  & \(150 \times 150 \times 96\)   & 9  & 1.6 billion & 16     &  $9.49 \times 10^{-4}$  & 400& 60 & \num{0.19+-0.003}           & \num{10.45+-0.053}                     \\
    \(10^{10}\) & \(250 \times 250 \times 128\)  & 9  & 5.8 billion & 13     &  $4.79 \times 10^{-4}$  & 200& 160 & \num{0.25+-0.005}           & \num{11.41+-0.110}                     \\
    \(10^{11}\) & \(500 \times 500 \times 256\)  & 5  & 8 billion & 11     &   $2.73 \times 10^{-4}$   & 100& 360 & \num{0.37+-0.004}           & \num{24.72+-0.457}           \\
    \(10^{12}\) & \(500 \times 500 \times 256\)  & 9  & 46.7 billion & 10     &   $2.02 \times 10^{-4}$  & 25 & 840 & \num{0.38+-0.011}           & \num{16.65+-0.448}                       \\
    \hline
    \end{tabular}
    \label{tab:sim_details}
  \end{center}
\end{table*}

For each spectral element grid listed under column $N_e$, the RBC simulation is first started with the lowest polynomial order $p=3$. Each case is initialized with zero velocity field and a linear temperature profile, $T = 1 - z$, along the vertical direction. Rayleigh number $Ra$ and polynomial order $p$ are incrementally ramped up as each run attains a statistically steady turbulent state. For some higher $Ra$ cases which use the same spectral element grid as a previous case ,but with increased polynomial order $p$, the solution dataset from the preceding case is used as initial condition for the new one thus allowing the simulations to attain statistical stationarity at a reduced computational expense. The averaging periods $\Delta t$ listed in \autoref{tab:sim_details} are taken in the respective steady-state regimes.

\subsection{Some physical results for $\Gamma=4$}
\begin{figure*}[!h]
\centering
\includegraphics[width=1.0\textwidth]{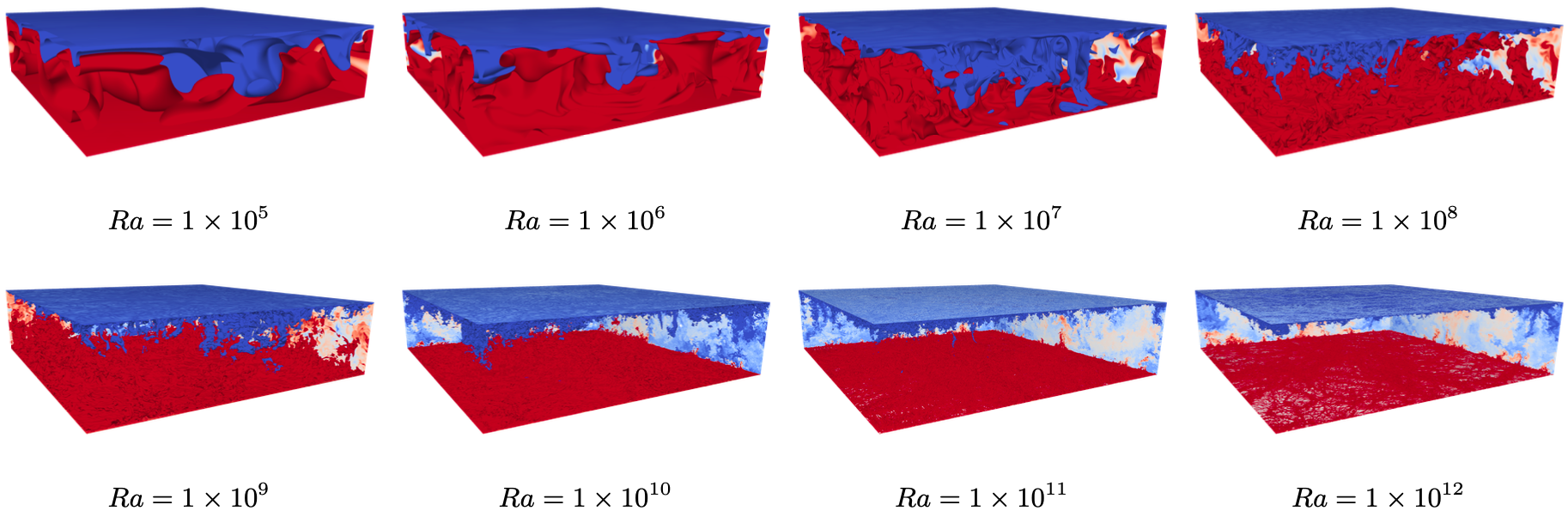}
\caption{Overview scenes for all Rayleigh number cases with $\Gamma=4$}
\label{fig:overview_3d}
\end{figure*}

\begin{figure*}[!h]
\centering
\includegraphics[width=1.0\textwidth]{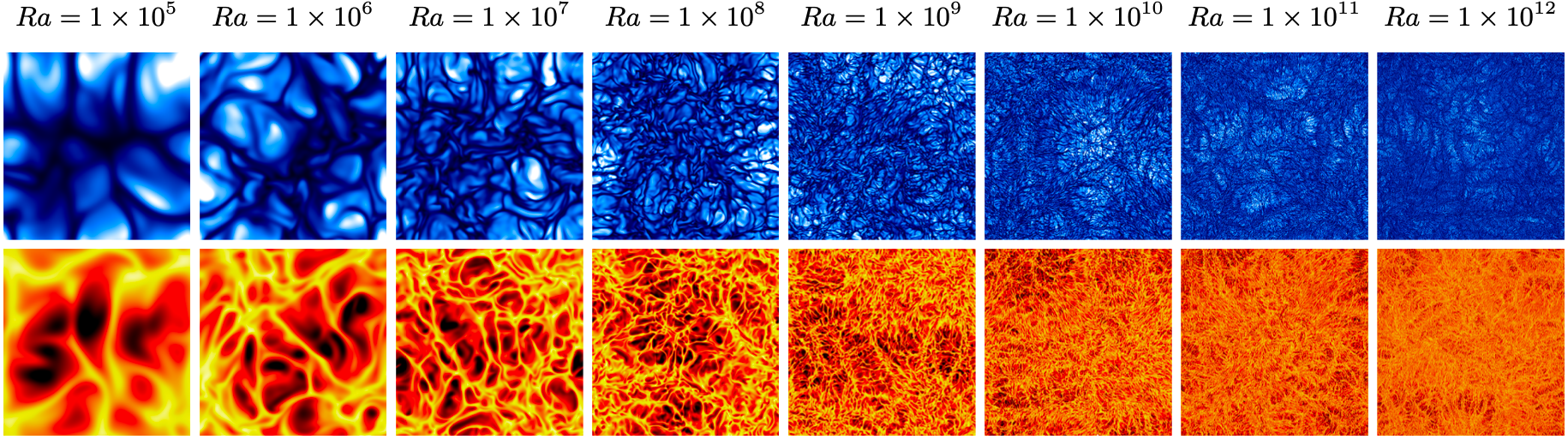}
\caption{Temperature contours in the top (top row) and bottom (bottom row) boundary layers for $\Gamma=4$}
\label{fig:bl_slices}
\end{figure*}

An instantaneous perspective scene of the statistically steady state regime is shown in \autoref{fig:overview_3d} for all Rayleigh number cases. It displays a temperature isosurface for hot (red) and cold (blue) isolevel, highlighting the thermal plume structure as well as temperature slices in the periodic directions. Additionally, slices through the thermal boundary layers at the top and bottom wall are shown in \autoref{fig:bl_slices} emphasizing the decreasing length scales and the resultant finer structures of the thermal plume network with increasing $Ra$.

\begin{figure}[t]
\centerline{\includegraphics[width=0.47\textwidth]{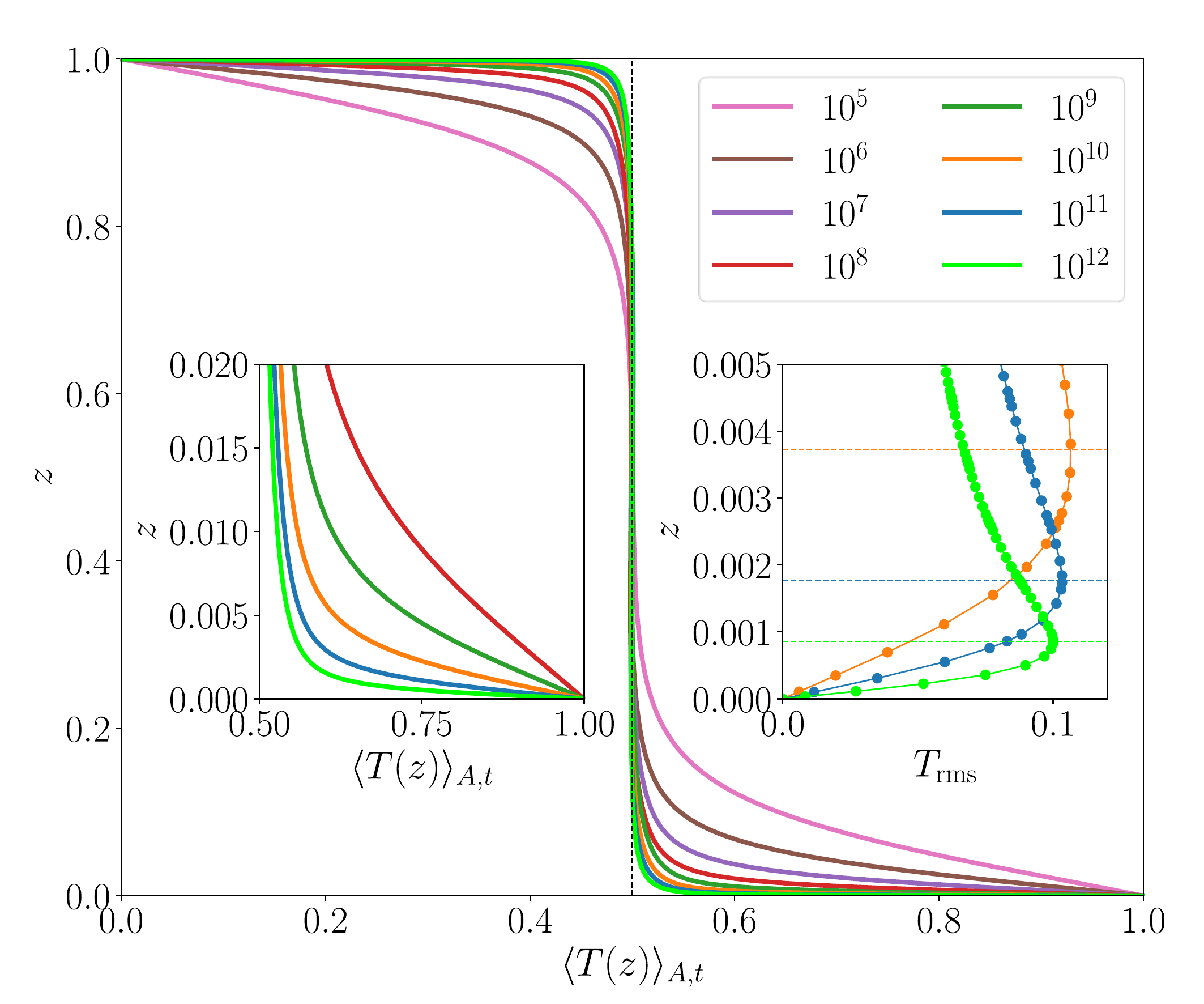}}
  \caption{
  Temperature profiles for $10^5 \leq Ra \leq 10^{12}$ and $\Gamma=4$.
  The left inset shows a close-up of the profiles near the bottom plate for the five highest values of $Ra$.
  The right inset shows the root-mean-square profiles of $T$, near the bottom wall for the three highest $Ra$.
  }
\label{fig:temp_prof}
\end{figure}

As mentioned in \autoref{ssec:case}, well-resolved DNS of RBC require the boundary layer to be captured with at least 10 collocation points. The mean and root-mean-square (rms) temperature profiles plotted in \autoref{fig:temp_prof} show that the simulations satisfy this requirement. The vertical profiles are obtained by averaging over both the horizontal plane and time. The height of the thermal boundary layer correspond to the maxima of the $T_\mathrm{rms}$ profiles. The markers on the $T_\mathrm{rms}$ profiles (right inset) show that even at the highest $Ra$, there are at least 10 collocation points in the thermal boundary layer.

Additionally, the profiles also highlight the shrinking height of the thermal boundary layers with increasing $Ra$. Correspondingly, the bulk region of the flow, where the mean temperature is about $(T_b+T_t)/2)=0.5$, expands to fill almost the full extent of the domain. This was also evident in \autoref{fig:overview_3d}, which shows the temperature isosurfaces corresponding to isolevels $T = 0.4$ and 0.6 for each case.

A key focus of numerous studies on RBC is the scaling of convective heat and momentum transport with Rayleigh number. The global convective heat transport is measured by the non-dimensional Nusselt number, $Nu$, which is the ratio between the total heat transfer in turbulent convection and that due to thermal conduction. The Nusselt number is measured as $Nu = 1 + \sqrt{Ra Pr}\langle u_z T \rangle_V$, where $\langle \cdot \rangle_V$ denotes a volume average, and $u_z$ and $T$ are the vertical component of velocity and temperature, respectively. The Reynolds number, $Re$, is defined as the ratio between inertial and viscous forces in the fluid. This dimensionless number quantifies the turbulent momentum transport across the layer and thus also the intensity of turbulence of the flow. It is $Re = UH/\nu$, where U is the RMS velocity, again computed as a volume average.

\begin{figure}
\centerline{\includegraphics[width=0.50\textwidth]{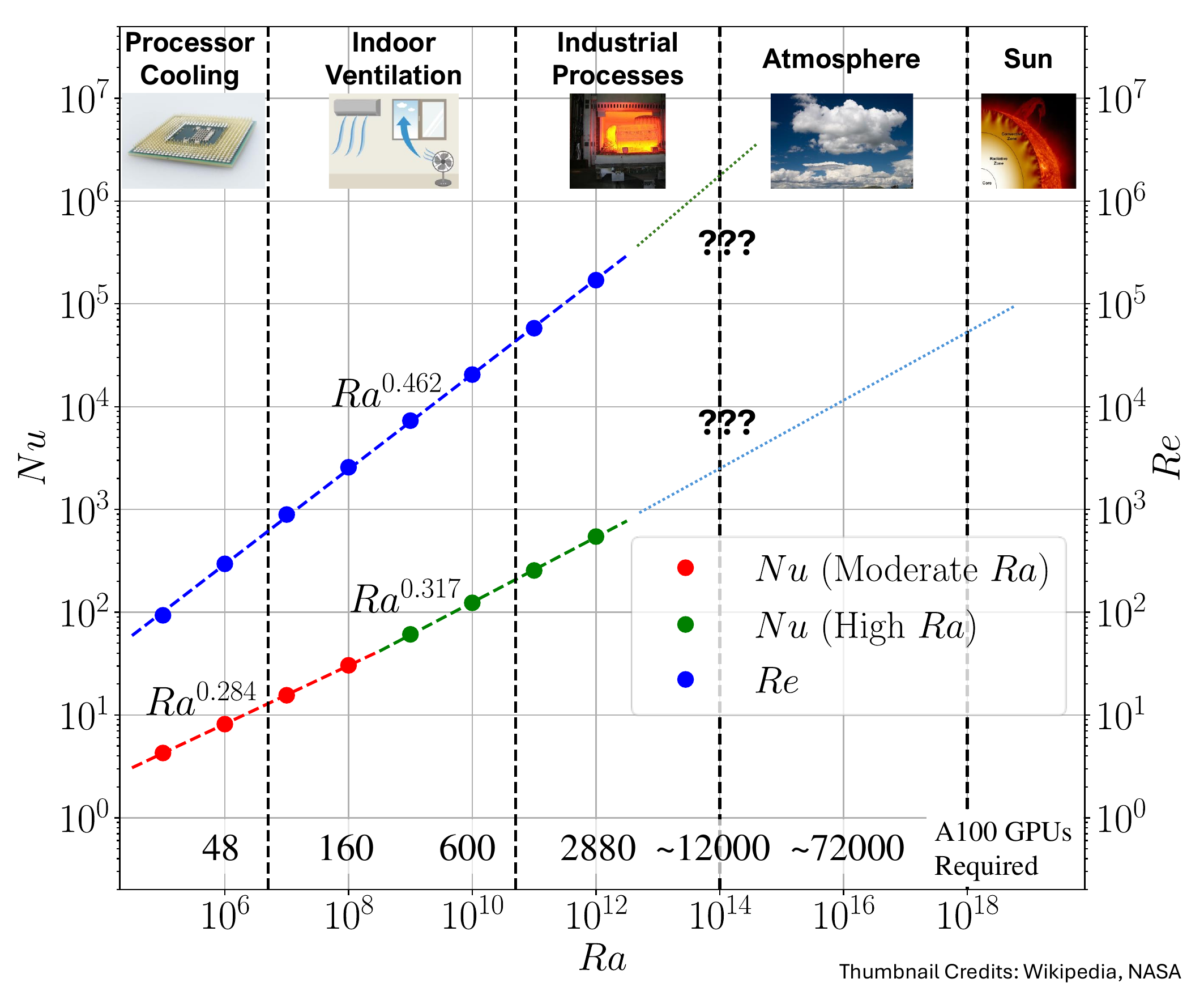}}
\caption{Scaling of the global quantities for heat transfer, $Nu$, and momentum transfer, $Re$ as functions of $Ra$. The relevant phenomena covered by the different ranges of $Ra$ are also indicated. The total number of A100 GPUs required for the simulations are estimated from the existing runs based on $N_{BL} \ge 10$}. 
\label{fig:nu_re_scale}
\end{figure}

The variation of both non-dimensional numbers, $Nu$ and $Re$, with respect to $Ra$ is of seminal importance to predict the heat transfers in a wide range of applications. In \autoref{fig:nu_re_scale}, both numbers are plotted against $Ra$ along with the corresponding scaling exponents of piecewise power law fits to the data (which we took for simplicity here). The scaling of $Nu$ is different at moderate and high $Ra$ and in agreement with classical theoretical predictions~\cite{Malkus:PRSA1954,Spiegel:ARAA1971}. However, it becomes also clear that many phenomena of interest require even larger simulations that go beyond the capacities of a systems such as JUWELS Booster and will only be accessible with exascale systems such as JUPITER. It has to be seen how this scaling proceeds to higher $Ra$ indicated by the question marks in the figure. 

\begin{figure}[h]
\centerline{\includegraphics[width=0.50\textwidth]{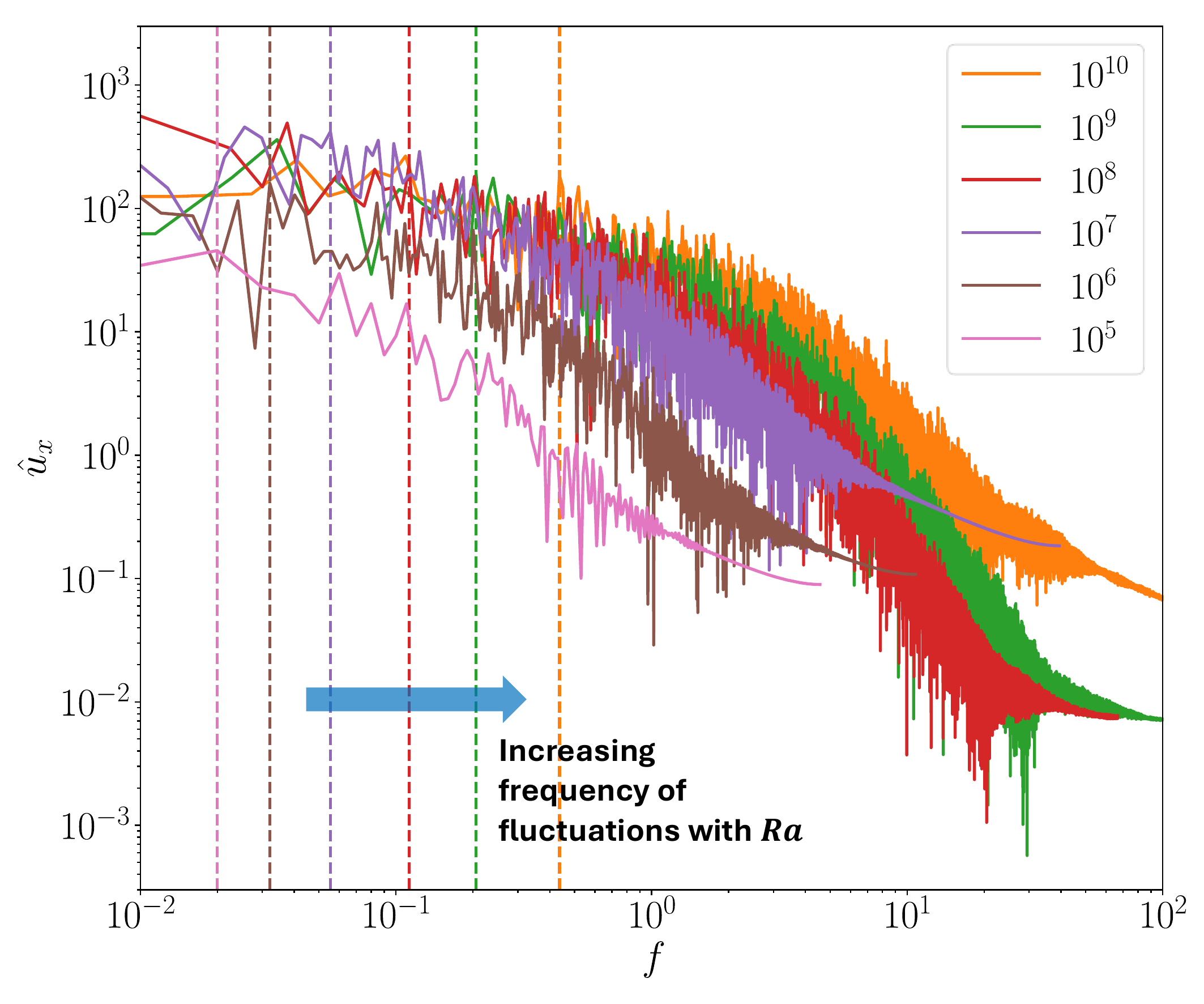}}
  \caption{
  Frequency spectra of velocity measured by a probe in the thermal boundary layer 
  }
\label{fig:probe_fft}
\end{figure}

To further understand the resulting flows and bolster the need for in-situ visualization of high $Ra$ simulations, the temporal fluctuations of the velocity field are examined. A probe was placed in the middle of the horizontal plane at the edge of the thermal boundary layer near the bottom plate to sample the velocity and temperature fields at every 10th time step. Fast Fourier Transforms were then used to compute the frequency spectra of the data collected from these probes for $10^5 \leq Ra \leq 10^{10}$ shown in \autoref{fig:probe_fft} along with the frequencies corresponding to the peak amplitudes, which progressively increase with $Ra$. Furthermore, the tails of the spectra also extend to higher frequencies with increasing $Ra$, indicating that the velocity fluctuations increase in frequency.

\begin{figure*}[!h]
\centering
\includegraphics[width=1.0\textwidth]{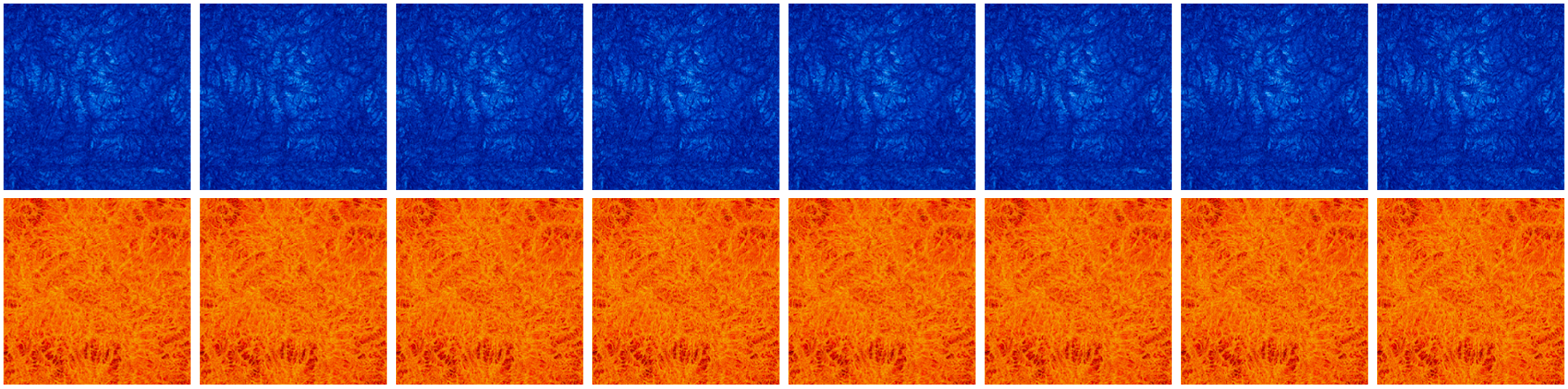}
\caption{Sequence of boundary layer snapshots at $Ra = 10^{12}$ written out each 100 time steps (from left to right). Fully resolved videos for the top and bottom boundary layer are added in the Supplemental Material of this article.}
\label{fig:bl_animate}
\end{figure*}

This is an interesting result both from the physics perspective as well as from the HPC viewpoint. Firstly, it indicates that the boundary layers become increasingly fluctuation-dominated as the $Ra$ increases~\cite{Samuel2024}. Secondly, this proves that as the $Ra$ increases, the solution field must be analysed over shorter time intervals to gather a concrete picture of the flow dynamics. Finally, \autoref{fig:bl_animate}, which shows a series of snapshots taken at free-fall time intervals of 0.02 units (every 100 time steps) for the $Ra = 10^{12}$ case, capture the flow smoothly with sufficient time-resolution.

Overall, the discussion makes it clear that RBC can be efficiently studied on GPU-accelerated supercomputers and can currently be achieved on JUWELS Booster $Ra=\num{e12}$ for $\Gamma=4$ with reasonable resolution. However, there is a clear physical motivation to overcome the current HPC limitations with respect to size and data handling in order to fully understand the dynamics in the boundary layers and to calculate flows with Rayleigh numbers in the range $10^{13}$ to $10^{15}$, which are theoretically possible with exascale systems.

\section{HPC performance results}
\label{sec:performance}
This section focuses on and analyzes the HPC performance of NekRS, the large-scale simulation at $Ra=10^{12}$ and the ASCENT workflow with respect to the additional cost of in-situ visualization in real applications at scale. The starting point of the evaluation is the real setup and the number of nodes specified by the memory requirements of the application and the desired time-to-solution. The perfect, i.\,e. maximally efficient, application regime of the pure ASCENT workflow is only discussed in passing. Four aspects are analyzed in four sections: The scaling properties of (pure) NekRS, the GPU performance of large runs, the overhead costs of in-situ visualization depending on the applied pipeline, and the different possible applications of ASCENT. Finally, the novelty is emphasized.

\subsection{Scaling of RBC on JUWELS Booster}
Before the actual analysis of the in-situ workflow, the weak and strong scaling properties of (pure) NekRS are first evaluated for the specific RBC case on JUWELS Booster. This is important because the analysis of RBC applications only makes sense if the cases are in the generally efficient scaling range. 

\begin{figure}
\centerline{\includegraphics[width=0.50\textwidth]{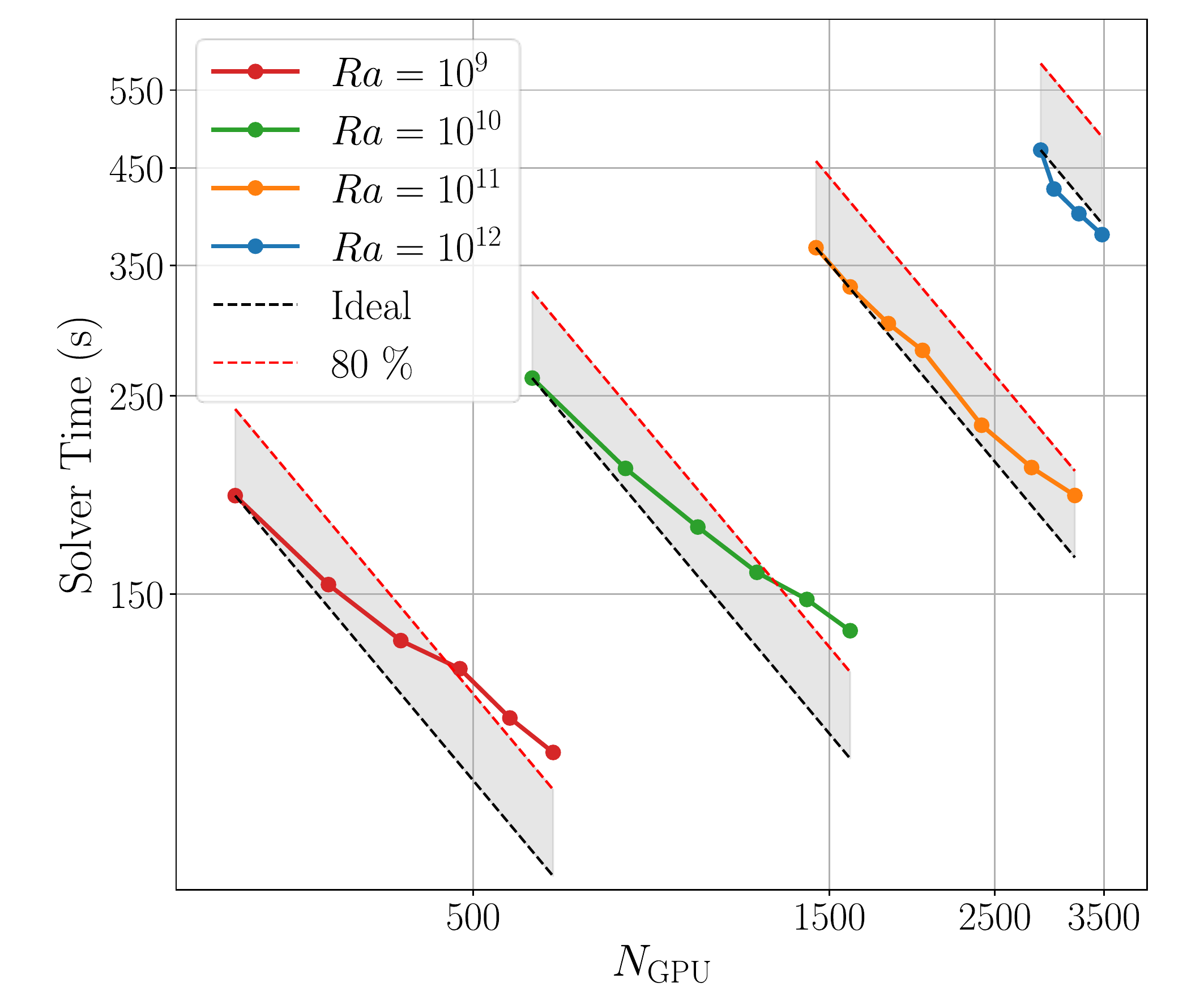}}
  \caption{
  Strong scaling of NekRS on JUWELS Booster for RBC at $Ra = 10^{9}$, $10^{10}$, $10^{11}$, and $10^{12}$.
  }
\label{fig:str_scale}
\end{figure}

The weak scaling was investigated for different polynomial orders with regard to increasing Rayleigh number and thus increasing element count (cf. \autoref{fig:nu_re_scale} and \autoref{tab:sim_details}). For this purpose, grids were built for the respective polynomial orders, which completely resolve the resulting flow of the respective Rayleigh number and are calculated on an (almost) minimum number of nodes in order to enable a maximum number of elements per GPU. The scaling was good across all nodecounts, but is not explicitly shown here. \autoref{fig:str_scale} shows for the cases with $Ra$ \num{e9} to \num{e12} the scaling in the strong-scale sense instead, i.\,e., with a decreasing number of elements per GPU. It is also good as long as a minimum number of elements per GPU is guaranteed (5400 elements/GPU for \num{e9} and 6500 elements/GPU for \num{e10}). For \num{e12}, the base point appears to be a negative outlier, which means that all other points have a parallel efficiency of over 100\,\%.

\subsection{Execution and performance of large-scale RBC simulation on JUWELS Booster}
Scaling is only the first prerequisite for successful HPC simulations. In practice, the time-to-solution is the relevant parameter, which depends on the general node performance and the algorithm used and can be reduced by using additional nodes if the scaling properties are good. In fact, scaling is typically better the slower the node performance of an application is, as this reduces the relative proportion of communication.

\begin{figure}
\centerline{\includegraphics[width=0.5\textwidth]{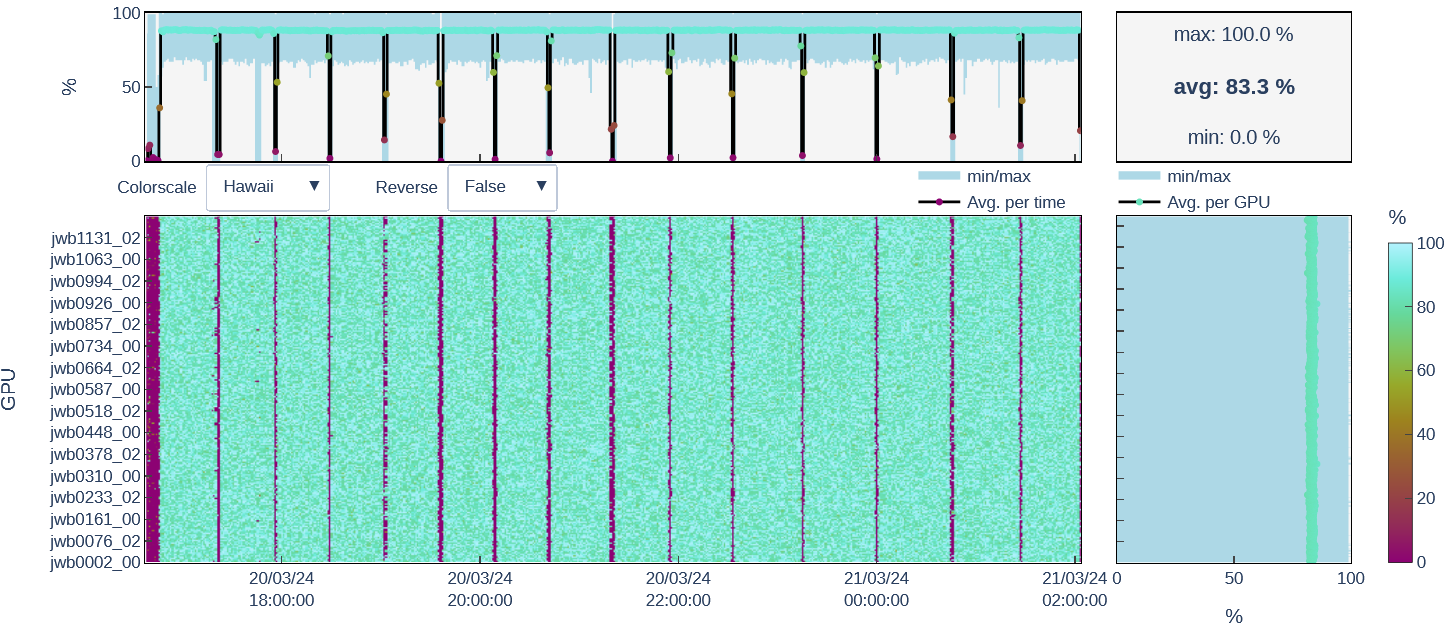}}
  \caption{
  GPU usage of the part of the large-scale simulation at $Ra=10^{12}$ without visualization
  }
\label{fig:gpu_perf}
\end{figure}

\autoref{fig:gpu_perf} shows the GPU usage of the part of the large-scale simulation without in-situ visualization in the form of an overview of the system monitoring (LLview\footnote{https://github.com/FZJ-JSC/LLview}). The average usage number of this job was 83.3\,\%. The periodic drop in GPU usage is due to the writing of checkpoint files and underlines once again that classic I/O is not suitable for creating time-resolved visualizations\footnote{Doubling the checkpoint file write interval would reduce GPU usage by approximately another 20 percentage points.}. In total, this part of the large-scale simulation required about \SI{7.5}{\hour} on 840 JUWELS booster nodes.

The 840 nodes/90\,\% JUWELS booster run was carried out with the help of SLURM reservations. This mechanism blocks other users from using the subset reserved for these runs. This reservation had 20 extra nodes, i.\,e., a total of 860 nodes were reserved, to account for possible node failures which can be in essence of two different kinds: First, hardware issues that might require manual intervention/repair; or second, transient software issues that are automatically cleared up after a certain time. This conservative 2\,\% buffer was chosen to prevent stalling the experiments queued up on the reservation, if just a handful of nodes end up having problems (transient or otherwise). Overall, several general system and code aspects were also learned during the execution of the large RBC run:
\begin{itemize}
\item Before each job start, a health check is run during the prologue. If the health check is not satisfactory in all the allocated nodes, the failed nodes are drained and the job is rescheduled for a later time. This works well for everyday jobs, but for very large jobs, individual nodes can critically get on the way to run the experiments, resulting in jobs being rescheduled many times, as the probability of a node failing increases with the size of the job. This highlights that running at scale requires a closer collaboration with system administration teams, and that current strategies employed by batch systems are not necessarily fine tuned for this case. In this line, a batch scheduler could support malleable scheduling of jobs, where for large cases a buffer $N$ of nodes is automatically added, which can be immediately used in case of a number of nodes $<N$ fail to clear the prologue. This is currently not supported in SLURM, and is anyway not without downsides, since that effectively could delay the start of jobs run without reservations and would overall decrease the system utilization.

\item With an increase in job sizes, hardware errors or problems become much more likely. Some of these issues might result in a crash. At present, JUWELS Booster does not support automatic re-queueing of failed jobs, except if the failure was triggered during the prologue. The reason behind is that it is often not trivial to automatically distinguish between a hardware error that is independent of the running job, and errors that might be triggered by the job and require human attention and cleanup before re-execution. As a consequence, large jobs must be monitored closely to avoid an idling machine during these large reservations. On the code side, this problem could at least be improved if a follow-up job is automatically submitted after each checkpoint file is written, which automatically continues from the last checkpoint file in the event of a hardware error. For the RBC simulations carried out here, chain jobs were submitted that automatically identify the last checkpoint file written and continue from there.

\item Flipping links -- short interruptions in the state of network links, in the order of tens of seconds -- occurred in several runs of the RBC case. These flips on the InfiniBand interfaces can be very disruptive for the jobs, and are a well-known problem on this architecture. To increase the chances on the software side that the simulation is not terminated in the event of link flip, the following UCX settings were also set:
\begin{lstlisting}
UCX_DC_MLX5_FC_ENABLE=y
UCX_DC_MLX5_TIMEOUT=10000000.00us
UCX_DC_MLX5_RNR_TIMEOUT=10000.00us
\end{lstlisting}
These are not considered a solution, but rather a workaround that increases the chances of successfully finishing an experiment, even in the presence of a flipping link and at the cost of a significant time overhead in affected time steps. The settings significantly reduced the occurrence of terminating flipping links for the RBC setup, but not completely. After the simulations presented here were completed, the system integrator identified the bus bars in the compute blades as the source of signal noise behind the flipping link events and proceeded to replace them on the full system, further reducing the occurrence of these events.
\end{itemize}

\subsection{In-situ pipeline evaluation}
A set of three images was used as a use case for the in-situ workflow, consisting of the overview scene from \autoref{fig:overview_3d} and the two slices through the upper and lower boundary layer shown in \autoref{fig:bl_slices}. The overview scene in ASCENT is composed of several partial images, two slices on the walls, two vertical slices in the background and an isocontour. 4096 pixels were used for all renderings in this work for the longest side. As the shorter side in the rectangular overview scene is only half the length, this results in 2048 pixels for this length. The simulations were monitored with JuMonC~\cite{witzler2024}.

\autoref{tab:sim_details} has already listed the average time per time step without and with in-situ visualization (incl. solution of the NSEs). The measurement always refers to the rendering of the overview scene and the two boundary layer slices. Interestingly, the in-situ time step is the most expensive for the $Ra=\num{e11}$ case. This is because this case has the same number of elements (but only with polynomial order 5 compared to 9) as the $Ra=\num{e12}$ case, which uses more than twice as many nodes. In general, the standard deviation increases significantly for larger cases and the largest absolute temporal deviation within the test sample was observed for the $Ra=\num{e11}$ case, which is not surprising due to the highest total write time. Measurements have also verified that time steps that do not make a visualization are on average as fast as time steps that were calculated with NekRS without ASCENT and thus do not generate any overhead as long as no ascent::execute is called in these time steps to read out the action trigger\footnote{Often, but not in all cases, it has been observed that the first (and sometimes second) step after an in-situ step is also significantly slower (up to 50\,\%). This is not critical for the overall overhead, but worth noting.}. Consequently, the total overhead of a simulation with in-situ visualization can be estimated as $(t_{vis}+(n_{vis}-1)*t_{novis})/(n_{vis}*t_{novis})$. Here $n_{vis}$ is the in-situ interval. 

Technically, it is not a problem not to always render every image. This means, for example, that it is possible to output the cut through the boundary layer on the warm wall every 100 time steps and the overview scene and the cut through on the cold side only every 1000 time steps. In this scenario, the in-situ visualization results in an overhead of 16.8\,\% for the simulation. If only one slice of a boundary layer were visualized every 100 time step, the overhead would be reduced to 14.1\,\%.
For the present large-scale simulation at $Ra=10^{12}$ rendering all three images with $n_{vis}=100$ this results in an overhead of 42.4\,\%. The costs of the individual parts of the visualization pipeline are shown systematically for two Rayleigh numbers in \autoref{tab:pipeline}. This is also evaluated for $Ra=\num{e9}$ on different node numbers. 

Some general conclusions can be drawn from the present performance study:
\begin{itemize}
\item The vertical slice is cheaper than the horizontal slice because the resulting number of elements is smaller.
\item The more monochrome a slice is, the cheaper it is to render.
\item Several similar slices increase the costs almost linearly from the second slice onwards. The theoretically better inclusion of more GPUs does not seem to bring any advantage.
\item The zoom factor, which determines how large the edges are in the final rendered image, has an influence on performance. The smaller the margins, the more expensive it becomes. When the margins are completely gone (this occurs between 1.6 and 1.7), it becomes cheaper again, as less data has to be rendered.
\item Sometimes quite high standard deviations occur, which are usually due to individual outlier values that take significantly longer to render.
\end{itemize}
\begin{table*}
  \caption{Evaluation of time (in seconds) taken by various parts of the in-situ visualization pipeline for the $Ra=\num{e12}$ case on 840 nodes as well as the $Ra=\num{e9}$ case computed on 60 and 100 nodes, respectively.}
  \begin{center}
    \begin{tabular}{lrrrr}
    \hline
    $Ra$ & \(10^{12}\)(840 nodes) & \(10^{9}\)(60 nodes) & \(10^{9}\)(100 nodes) \\
    \hline
    \(t_{novis}\) & \num{0.38+-0.011} & \num{0.19+-0.003} & \num{0.13+-0.005} \\
    \(t_{vis}\) & \num{16.65+-0.448} & \num{10.45+-0.053} & \num{9.80+-0.054} \\
    \hline    
    \(t_{contour}\) & \num{5.12+-0.014} & \num{2.32+-0.011} & \num{2.10+-0.003} \\
    \(t_{vertical}\) & \num{4.07+-0.067} & \num{2.00+-0.010} & \num{1.86+-0.014} \\
    \(t_{horiz.}\) & \num{4.14+-0.031} & \num{2.05+-0.041} & \num{1.89+-0.013} \\
    \hline    
    \(t_{1xslice}\) & \num{5.57+-0.022} & \num{3.42+-0.020} & \num{3.43+-0.018} \\
    \(t_{2xslices}\) & \num{8.41+-0.021} & \num{6.00+-0.025} & \num{6.24+-0.027} \\
    \(t_{4xslices}\) & \num{14.15+-0.053} & \num{11.57+-0.841} & \num{11.90+-0.047} \\
    \(t_{8xslices}\) & \num{25.90+-0.462} & \num{23.55+-1.731} & \num{23.29+-0.242} \\
    \hline    
    \(t_{zoom1.2}\) & \num{5.62+-0.041} & \num{3.31+-0.015} & \num{3.37+-0.021} \\
    \(t_{zoom1.5}\) & \num{5.74+-0.038} & \num{3.40+-0.025} & \num{3.41+-0.026} \\
    \(t_{zoom1.6}\) & \num{5.77+-0.032} & \num{3.40+-0.022} & \num{3.43+-0.030} \\
    \(t_{zoom2.0}\) & \num{5.78+-0.026} & \num{3.39+-0.016} & \num{3.40+-0.012} \\
    \(t_{zoom5.0}\) & \num{5.67+-0.036} & \num{3.44+-0.023} & \num{3.33+-0.013} \\
    \hline
    \end{tabular}
    \label{tab:pipeline}
  \end{center}
\end{table*}

The results of the investigation of the scaling capability of the $Ra=\num{e9}$ case are shown in \autoref{tab:ascent_scal}. The scaling of the in-situ step is definitely less efficient than NekRS in general. However, this is not surprising, since slices, for example, only involve a limited number of GPUs in individual visualization steps and thus an increasing number of GPUs are ideling. 
\begin{table}
  \caption{Scaling of time steps with and without in-situ visualization of the full setup for the $Ra=\num{e9}$ case on different number of nodes.}
  \begin{center}
    \begin{tabular}{ccccc}
    \hline
    \(N_{nodes}\)      &       \(t_{novis} (s)\)   & \(e_{novis}\)     &       \(t_{vis} (s)\)   & \(e_{vis}\) \\
    \hline
    60 & \num{0.19+-0.003} & - & \num{10.54+-0.052} & - \\
    80 & \num{0.15+-0.003} & \num{0.93} & \num{9.76+-0.050} & \num{0.81} \\
    100 & \num{0.13+-0.005} & \num{0.90} & \num{9.44+-0.066} & \num{0.67} \\
    120 & \num{0.13+-0.003} & \num{0.76} & \num{9.77+-0.257} & \num{0.54} \\
    140 & \num{0.11+-0.003} & \num{0.75} & \num{9.67+-0.785} & \num{0.46} \\
    160 & \num{0.11+-0.005} & \num{0.66} & \num{9.37+-0.036} & \num{0.42} \\
    \hline
    \end{tabular}
    \label{tab:ascent_scal}
  \end{center}
\end{table}

\begin{figure}
\centering
\begin{subfigure}[t]{0.5\textwidth}
    \centering
    \includegraphics[width=\textwidth]{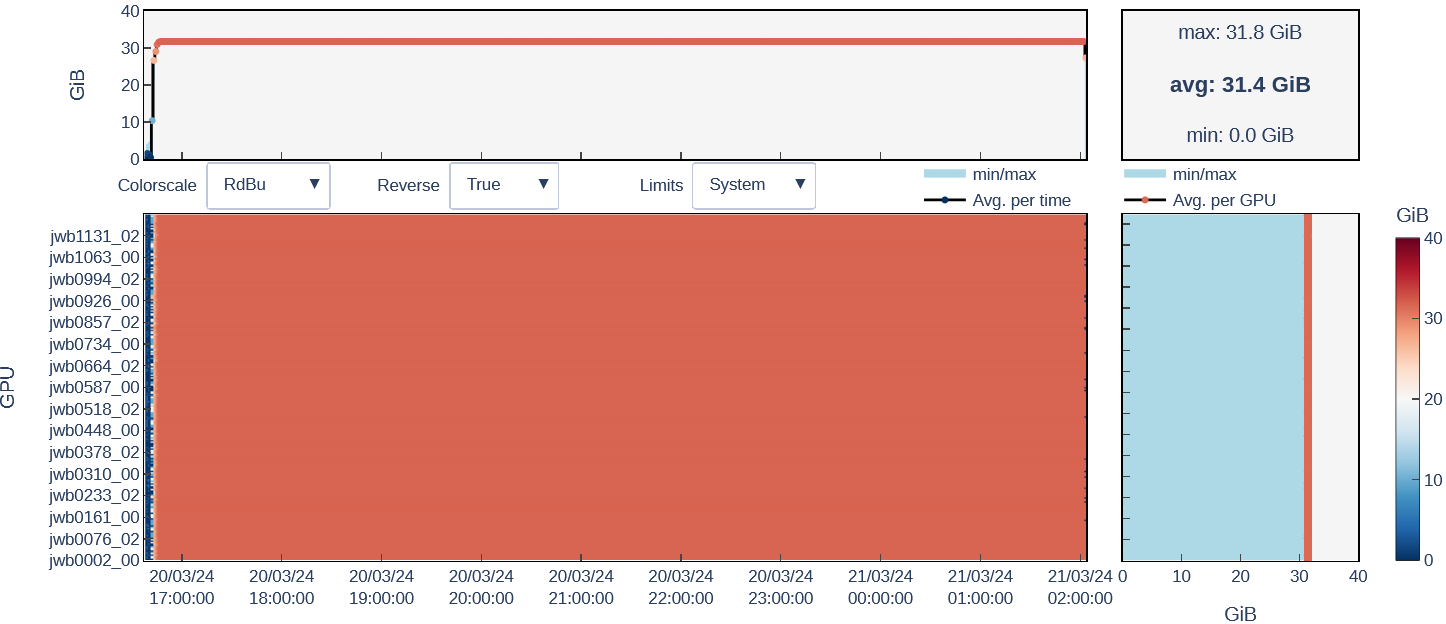}
\end{subfigure}

\begin{subfigure}[t]{0.5\textwidth}
    \centering
    \includegraphics[width=\textwidth]{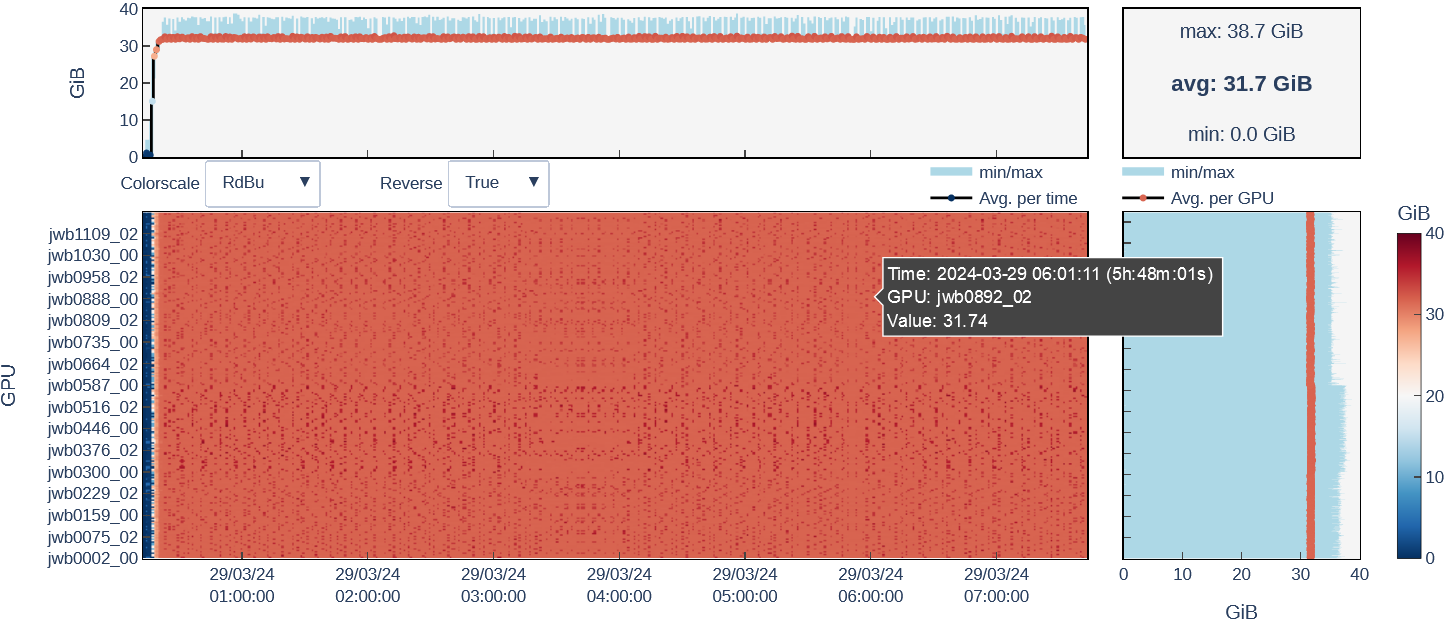}
\end{subfigure}

\caption{
GPU memory usage for a simulation without (top) and with in-situ visualization on 840 nodes
}
\label{fig:mem_usage}
\end{figure}

GPU memory usage is another interesting aspect of the in-situ workflow. \autoref{fig:mem_usage} compares this for a run without and a run with in-situ visualization. The average memory usage is approximately the same for both cases. However, the maximum memory usage of the run with in-situ visualization is significantly higher. This must be taken into account when setting up the run. In general, the memory usage in the in-situ case is significantly more variable, as can be seen both in the "speckled" red area and at the bottom right.

\subsection{Further analysis pipeline}
\label{ssec:fur}
The project aimed not only to leverage the capabilities of ASCENT for in-situ analysis and visualization but also to showcase a novel workflow facilitating collaboration across distinct supercomputing centers. A common scenario is that steady state data, such as the result of the present large-scale simulation on JUWELS Booster, is transferred to other supercomputing centers for further analysis. Sometimes the various cases with different $Ra$ are also executed on different supercomputers, because research teams are granted resources on several supercomputers. To ensure that the workflow developed in the course of this work is independent of the supercomputer used and works in general, the RBC simulations including the visualization were also carried out on Leonardo Booster, MareNostrum 5 ACC, and Polaris on a test basis. It was possible to reproduce the visualizations shown in the course of this work on all systems, but the performance on the additional supercomputers was not systematically evaluated.

To test whether the in-situ workflow also works with other CFD simulations using NekRS and to identify hidden case-specific implementations, the workflow was also used with reactive flows. For this purpose, NekRS together with the chemistry plugin NekCRF~\cite{kerkemeier2024} was applied to hydrogen and ammonia flames, also on up to 3360 GPUs. No general technical difficulties were identified and the results are currently undergoing quantitative analysis.

In the context of exascale systems, such as JUPITER with its 288 Grace cores per node, it can also be interesting to copy the data to the CPU despite the overhead. On the one hand, the GPUs in the compute nodes may not support ray tracing, for example, as they tend to be optimized for training AI models. On the other hand, this could be a way to use the host CPU cores effectively while the GPUs calculate the actual simulation. For the RBC use case in this work, a write interval of 100 means that for the $Ra=\num{e12}$ case it is approximately \SI{38}{\second} between two write instances. Depending on the rendering effort, copying may be worthwhile here if the CPUs are otherwise not used at all. This is currently being analyzed quantitatively in further work.

\subsection{Novelty}
From a physical point of view, the novelty of this paper is obvious. The presented workflow is the first to enable simulations at a very high Rayleigh number of \(10^{12}\) with \(\Gamma=4\) and capturing the full time-resolved visualization and analysis of the turbulent flow dynamics. Data of this caliber, providing unprecedented details of the complex turbulent heat transfer mechanisms in a very turbulent regime, has never before been attainable or published. The fully time-resolved visualizations paves the way to a deeper understanding of turbulent heat transfer processes under such extreme conditions which cannot be obtained from laboratory experiments.

The HPC-novelty of this work is twofold: First, this is the first time that NekRS has been coupled with ASCENT and demonstrated scalability at such large runs. Second, this is an implementation, usage, and demonstration of zero-copy GPU-to-GPU in situ visualization in a real-world scenario. Moreover,  a different application scenario is addressed that demands a very high visualization frequency, introducing unique HPC-challenges. The implementation enabled the largest CFD simulations on JUWELS Booster and is thus a very important step towards the first European CFD exascale simulations on JUPITER. It differs from the established US system (e.\,g., no Slingshot). This is also illustrated by the described challenges such as flipping links, which had to be addressed. In general, the new workflow enables high-frequency visualization at a manageable computational cost. The concrete implementation is coupled to NekRS, but can be applied to other codes right away. To summarize, numbers for the visualization for four dimensionless time units (``free fall times'') are given: Visualizing the case of $Ra=\num{e12}$ using classical post-processing methods implies a handling of $200*1.2~\mathrm{TB}=240~\mathrm{TB}$ data over four hours of computing on 3360 GPUs (without I/O). Conventional visualization would then potentially consume hours or even days. The in-situ approach removes these bottlenecks, eradicating the massive data movement. By circumventing the need to write out the full $240~\mathrm{TB}$ dataset, the presented workflow extends the simulation timescale by a mere $40~\mathrm{minutes}$ - a fractional overhead compared to the crushing data burden of traditional methods. This paper demonstrates the first successful implementation of such a transformative in-situ workflow, eradicating previous constrains, and enabling scientific visualization capabilities at unprecedented rates.

\section{Conclusions}
The focus of this work is an RBC setup with different Rayleigh numbers that enables deeper physical insights into turbulent heat transfer processes which are ubiquitous in many natural and technical flows. The direct numerical simulation data shows that fluctuations in the viscous and thermal boundary layers play a central role in this dynamics. In these boundary layers, the structures are formed which carry the supplied heat locally into the bulk of the layer and drive the hard turbulence. Resolving their dynamics and structure by in-situ visualizations is thus essential for a deeper understanding of the system as a whole. The largest Rayleigh number case was calculated on 3360 GPU of JUWELS Booster, i.\,e., 90\,\% of the system, for almost 24 hours to a statistically steady state and then continued for 4 convective free-fall time units to visualize in particular the dynamical processes in the warmer boundary layer at the bottom by means of in-situ visualization completely time-resolved. An ASCENT in-situ workflow made this possible and resulted in an overhead of 16.8\,\%. This would have been impossible using classic checkpointing (individual checkpoint files take over 10 minutes). Overall, this was one of the largest single simulations ever performed on JUWELS Booster.

The integration of ASCENT into NekRS facilitated a seamless and efficient method for establishing in-situ visualization and analysis pipelines, eliminating the necessity for code recompilation. The performance was quantified using measurements on JUWELS Booster. Overall, this work is a good example of how modern HPC workflows, such as in-situ visualizations, enable deeper physical insights into the full three-dimensional dynamics in complex multi-scale systems, such as turbulent convection flows at very high Rayleigh numbers.

\section*{Acknowledgment}
This research used resources of the Argonne Leadership Computing Facility, a U.S. Department of Energy (DOE) Office of Science user facility at Argonne National Laboratory and is based on research supported by the U.S. DOE Office of Science-Advanced Scientific Computing Research Program, under Contract No. DE-AC02-06CH11357. This work was supported by Northern Illinois University. The authors acknowledge computing time grants for the project TurbulenceSL by the JARA-HPC Vergabegremium provided on the JARA-HPC Partition part of the supercomputer JURECA at Jülich Supercomputing Centre, Forschungszentrum Jülich, the Gauss Centre for Supercomputing e.V. (www.gauss-centre.eu) for funding this project by providing computing time on the GCS Supercomputer JUWELS at Jülich Supercomputing Centre (JSC), EuroHPC Joint Undertaking for proving compute time on Leonardo Booster (EHPC-EXT-2024E01-071), and funding from Innovative Algorithms for Applications on European Exascale Supercomputers (Inno4Scale, grant agreement No. 101118139) program. Support by the Joint Laboratory for Extreme Scale Computing (JLESC, https://jlesc.github.io/) for traveling is acknowledged. Furthermore, this work was funded by the European Union (ERC, MesoComp, 101052786). Views and opinions expressed are however those of the authors only and do not necessarily reflect those of the European Union or the European Research Council.

\bibliographystyle{elsarticle-num}
\bibliography{nekX}

\end{document}